\documentclass[aps,prd,twocolumn,amsmath,amssymb,showpacs,floatfix,nofootinbib]{revtex4}

\usepackage{amsmath}
\usepackage{hyperref}
\usepackage{graphicx, calc}
\usepackage{xcolor}
\usepackage{hhline}
\usepackage{cancel}
\usepackage[normalem]{ulem}
\usepackage{amssymb}
\usepackage{amsfonts}
\usepackage{enumerate}
\usepackage{afterpage}
\usepackage{color}
\usepackage{ae,aecompl}
\usepackage{soul}

\usepackage{multirow}

\usepackage[T1]{fontenc}
\usepackage{lmodern}

\newcommand{\be}{\begin{equation}}
\newcommand{\ee}{\end{equation}}

\newlength{\depthofsumsign}
\newcommand{\nsum}[1][1.4]{
    \mathop{%
        \raisebox
            {-#1\depthofsumsign+1\depthofsumsign}
            {\scalebox
                {#1}
                {$\displaystyle\sum$}%
            }
    }
}

\definecolor{darkred}{rgb}{0.8,0,0}

\begin{document}

\title[Biases from neutrino bias]{Biases from neutrino bias: to worry or not to worry?}
\author{Alvise Raccanelli$^{\star\let\thefootnote\relax\footnote{$^\star$Marie Sk\l{}odowska-Curie fellow},1}$, Licia Verde$^{1,2}$, Francisco Villaescusa-Navarro$^3$} 

\affiliation{
$^1$Institut de Ci\`encies del Cosmos (ICCUB), Universitat de Barcelona (IEEC-UB), Mart\'{i} Franqu\`es 1, E08028 Barcelona, Spain \\
$^2$ICREA, Pg. Lluis Companys 23, 08010 Barcelona, Spain \\
$^3$Center for Computational Astrophysics, 160 5th Avenue, New York, NY, 10010, USA
}

\begin{abstract}
The relation between the halo field and the matter fluctuations (halo bias), in the presence of massive neutrinos depends on the total neutrino mass; massive neutrinos introduce an additional scale-dependence of the bias which is usually neglected in cosmological analyses.  We investigate the  magnitude of the systematic effect on interesting cosmological parameters induced by neglecting this scale dependence, finding that while it is not a problem for current surveys,  it is non-negligible for future, denser or deeper ones depending on the neutrino mass, the maximum scale used for the analyses and the details of the nuisance parameters considered. However there is a simple recipe to account for the bulk of the effect  as to make it fully negligible, which we illustrate and advocate should be included in analysis of forthcoming large-scale structure  surveys. 
\end{abstract}

\maketitle

\section{Introduction}
\label{sec:introduction}
The clustering of large scale structure (LSS) is one of the key cosmological  observables, encoding precious information about the  Universe's content, the properties of its components and the nature of the primordial perturbations, which are highly complementary  to those that can be extracted from observations of the Cosmic Microwave Background (CMB). In particular the detailed shape of the matter power spectrum is crucial to both constrain the shape of the primordial power spectrum (and thus glean information about the mechanism that set out the initial conditions) and measure the  absolute neutrino mass scale.
While in the standard $\Lambda$CDM (and in the standard model for particle physics) neutrinos are assumed massless,  neutrino oscillation experiments indicate that neutrinos have non-zero mass see~\cite{1969GP} and  e.g.~\cite{GM08} for a review. However only the (square) mass splitting between the different mass eigenstates has been measured, not the absolute masses. Cosmological observations are key to constrain the sum of neutrino masses, $M_{\nu}$,  since free streaming  neutrinos can suppress clustering at small scales e.g., \cite{1980SvAL....6..252D,Huetal1998} and see~\cite{Lesgourgues:15} for a thorough  review. In particular, massive neutrinos streaming out from an over-density reduce the amount of matter  accumulating gravitationally  in that over density, this leads to a suppression of power below the free streaming scale and thus a scale-dependent growth rate of  perturbations \cite{eisensteinandhu}. Several works have quantified this both in the linear and non-linear regime analytically and with N-body simulations e.g.~\cite{Takadaetal2006, Kiakotouetal2008, Blasetal2014,Brandbyge:2010ge, Birdetal2012} and refs therein.  This effect makes it possible to place constraints on the sum of neutrino masses that is not accessible to future laboratory experiments~\cite{Osipowicz,Monreal,Doe}.
Recent cosmological analyses provide robust 95\% confidence upper limits $M_{\nu}<0.12$ eV \cite{Palanque-Delabrouille:2015}, $M_{\nu}<0.13$ eV \cite{Cuesta2016}, which, in combination with constraints from the mass splitting (e.g.~\cite{Bergstrom15, 2017JHEP...01..087E}),  implies that $0.058< M_{\nu}<0.13$ eV. While the lower limit is cosmology-independent (but assumes no massive sterile neutrinos) the upper limit is cosmology dependent and  would relax for models more complex than a  6+1 cosmological parameters  model (a standard $\Lambda$CDM with massive neutrinos) see e.g., \cite{Planck_2015}. 
This implies that future LSS surveys have the statistical power to detect the signature of non-zero neutrino mass if systematic effects can be kept under control e.g.~\cite{Carbone:2010ik,Hamann:2012fe,  Audrenetal2013, paco_15}.
Chief among these is the relationship between the clustering of dark matter and of its observable tracers (bias). Since tracers are hosted in dark matter halos,   one of the crucial ingredients to understand bias is to model correctly the bias of the halo field or halo bias. 

The large thermal velocities of cosmic neutrinos avoid their clustering within dark matter halos~\cite{Singh_2003, Ringwald_2004,Brandbyge_2010, paco_2011,paco_2012}. Dark matter halos abundance can thus be derived from the statistical properties of the CDM+baryons field, see~\citep{Ichiki-Takada, Castorina:2013wga, LoVerde_2014, Castorina:2015bma}. In cosmologies with massive neutrinos the CDM+baryon field is not only responsible for the abundance of dark matter halos, but also for their clustering \cite{paco14,Castorina:2013wga, loverde}. Since the CDM+baryons power spectrum differs in amplitude and shape with respect to the total matter power spectrum, defining halo bias with respect to total matter will induce a scale-dependent bias even on large-scales. In~\cite{paco14,Castorina:2013wga} it was shown that on large-scales in cosmologies with massive neutrinos the halo bias become scale-independent and universal if it is defined as:
\begin{equation}
b_{\rm cc}(k)=\sqrt{P_{\rm hh}(k)/P_{\rm cc}(k)} \, ,
\end{equation}
where $P_{\rm hh}(k)$ is the halo power spectrum and $P_{\rm cc}(k)$ is the CDM+baryons power spectrum, that can be estimated from the different transfer functions as:
\begin{equation}
P_{\rm cc}(k)=P_{\rm m}(k)\left(\frac{T_{\rm cb}(k)}{T_{\rm m}(k)}\right)^2 \, ,
\end{equation}
with:
\begin{equation}
T_{\rm cb}(k)=\frac{\Omega_{\rm cdm}T_{\rm cdm}(k)+\Omega_{\rm b}T_{\rm b}(k)}{\Omega_{\rm cdm}+\Omega_{\rm b}} \, ,
\end{equation}
where $T_{\rm cdm}(k)$, $T_{\rm b}(k)$ and $T_{\rm m}(k)$ are the transfer functions of CDM, baryons and total matter while $P_{\rm m}(k)$ is the total matter power spectrum. Here and hereafter unless otherwise stated there is an implicit redshift dependence in quantities such as bias and power spectra.

While even in CDM-only cosmologies the  constant linear bias model is incorrect and too simplistic to be employed unchecked in the analysis of  present and future surveys,  the extra scale dependence introduced by neutrino masses is usually neglected, but  it is important  to quantify the applicability of this approximation.

We  therefore investigate the systematic error (shift) introduced by neglecting this extra scale-dependence induced by neutrino masses on the best-fit of relevant cosmological parameters measured from galaxy clustering. It is timely to consider this observable, as it represents the main quantity to be measured by future cosmological galaxy surveys, and it has already been used to set low-z constraints on e.g. models of gravity, primordial non-Gaussianity, neutrino mass and standard model parameters (see e.g.~\cite{Samushia:2012, Raccanelli:2013growth, Zhao:2012, Ross:2013, Anderson:2013, Mueller:2016, Zhao:2017, BOSS:DR13}).
We concentrate on masses in the range above, but also explore higher masses both  for illustrative purposes and because in practice  when the analysis explores parameter space, higher masses might have to be considered.

This paper is structured as follows. In Section~\ref{sec:methodology}   we introduce the suite of simulations used to characterise the  $M_{\nu}$-induced scale dependence of the halo bias, our modeling of the observable, the full shape of the halo power spectrum, and the Fisher matrix-based approach to estimate the systematic shifts induced by neglecting the above scale dependence. In Section~\ref{sec:results} we present our results that are the forecasted shifts for a suite of forthcoming and future surveys. We also propose a simple and inexpensive way to  satisfactorily account for the effect. We conclude in Section~\ref{sec:conclusions}. The Appendix reports useful fitting formulae for the scale dependence of the halo bias.

\section{Methodology}
\label{sec:methodology}
\subsection{N-body simulations of halo bias  with massive neutrinos}
\label{sec:bias}
We use a subset of the \textsc{HADES} simulations \cite{HADES}, that we describe briefly here. Simulations have been run using the TreePM code \textsc{Gadget-III}, last described in \cite{Springel_2005}. Our simulations follow the evolution of $1024^3$ CDM plus $1024^3$ neutrino particles (only in models with massive neutrinos) in a periodic box of 1000 $h^{-1}$Mpc. The softening length of both CDM and neutrinos is set to $1/40$ the mean inter-particle distance. 
Initial conditions are generated at $z=99$, taking into account the scale-dependent growth factor/rate present in cosmologies with massive neutrinos employing the procedure outlined in \cite{Zennaro_2016}, using the Zel'dovich approximation. Massive neutrinos are simulated as a pressureless and collisionless fluid, i.e. using the particle-based method \cite{Brandbyge_2008, Viel_2010, Wagner12}, and we assume that neutrinos have degenerate masses. We consider two different cosmological models with $\sum m_\nu=M_\nu={0.0, 0.15}$ eV. The value of the other cosmological parameters are $\Omega_{\rm m}=\Omega_{\rm cdm}+\Omega_{\rm b}+\Omega_\nu=0.3175$, $\Omega_{\rm b}=0.049$, $n_s=0.9624$, $h=0.67$, $A_s=2\times10^{-9}$, in agreement with the results from Planck \cite{Planck_2015}. For each cosmological model we run 15 realizations with different random initial seeds and store snapshots at redshifts 0, 0.5, 1, 2 and 3.

\begin{figure}[htb]
\includegraphics[width=\columnwidth]{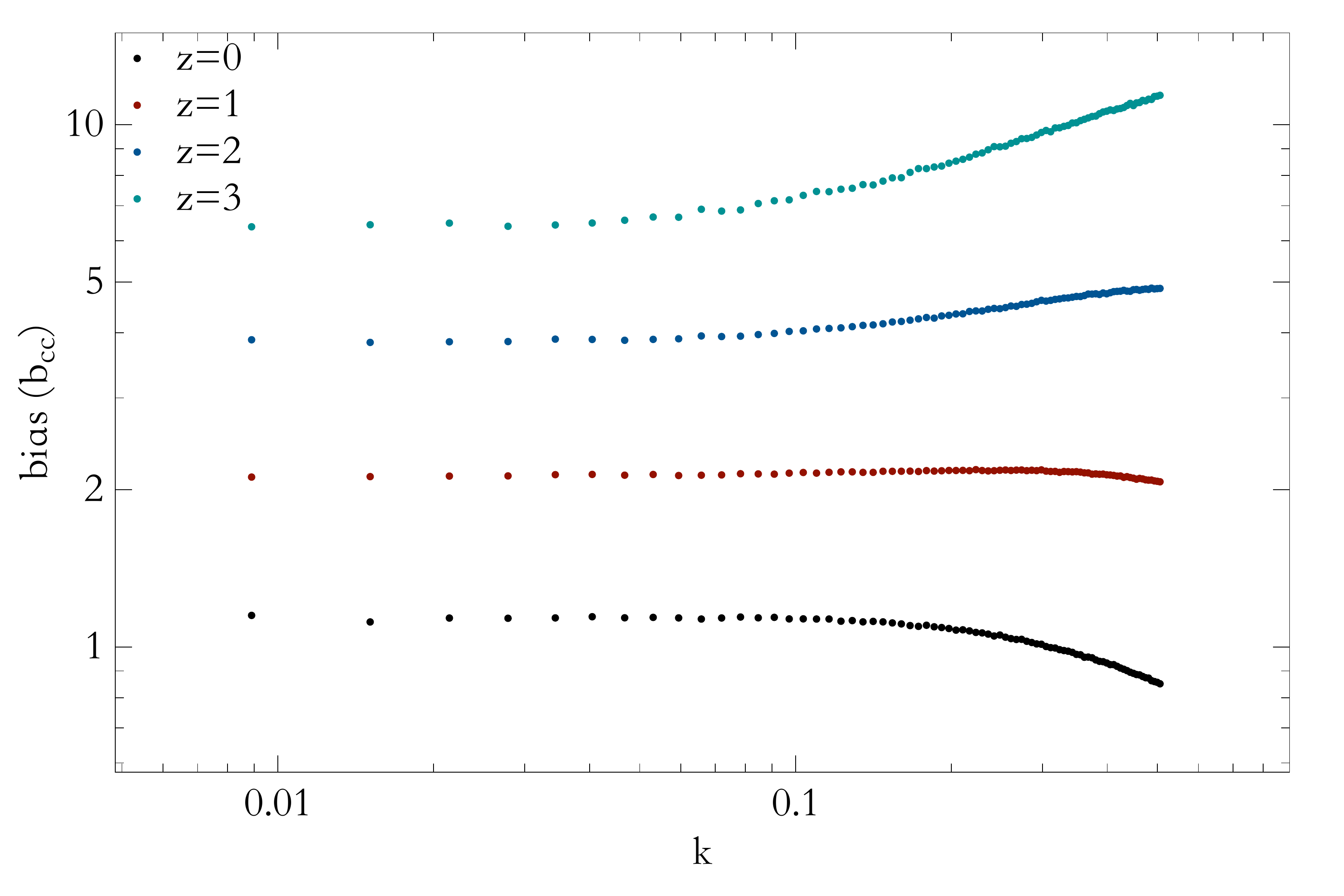}
\caption{Bias with respect to the CDM $b_{\rm cc}$ obtained from simulations, for different redshifts. The scale dependence of  $b_{\rm cc}$ does not depend on neutrino mass. For more details see text.} 
\label{fig:bias_cc_zi}
\end{figure}

To compute the halo bias we first estimate the matter and halo power spectra by  computing the density using the cloud-in-cell scheme and then the standard Fast Fourier transform procedure.  The bias is then obtained from the power spectra ratio, and we compute the mean and dispersion from the different realizations.  While in principle the cross (halo-matter) power spectrum  should be used to estimate the bias as $b_{\rm hm}=P_{\rm hm}/P_{\rm mm}$ in a way that is less sensitive to stochasticity, for our application, which results  we envision  should be applied to the analysis of galaxy surveys, the interesting quantity  actually depends on $P_{\rm mm}$, and is actually $b_{\rm mm}=\sqrt{P_{\rm hh}/P_{\rm mm}}$. As we discussed before, the clustering properties of dark matter halos in cosmologies with massive neutrinos will be controlled by the statistical properties of the CDM+baryons density field rather than the total matter density field. Thus, beyond $b_{\rm mm}(k)$ we will also compute $b_{\rm cc}(k)$ from the simulations.
In Figure~\ref{fig:bias_cc_zi} we show the bias with respect to the CDM $b_{\rm cc}(k)$ as obtained from the simulations, for halos with masses larger than $4\times 10^{12} M_\odot$; this is further explored in Appendix~\ref{sec:appendix}, where we also present our adopted smooth fitting formula for it, and in Figure~\ref{fig:bias_appendix} we show the comparison between the fit and the bias from the simulations.

\begin{figure}[htb]
\includegraphics[width=\columnwidth]{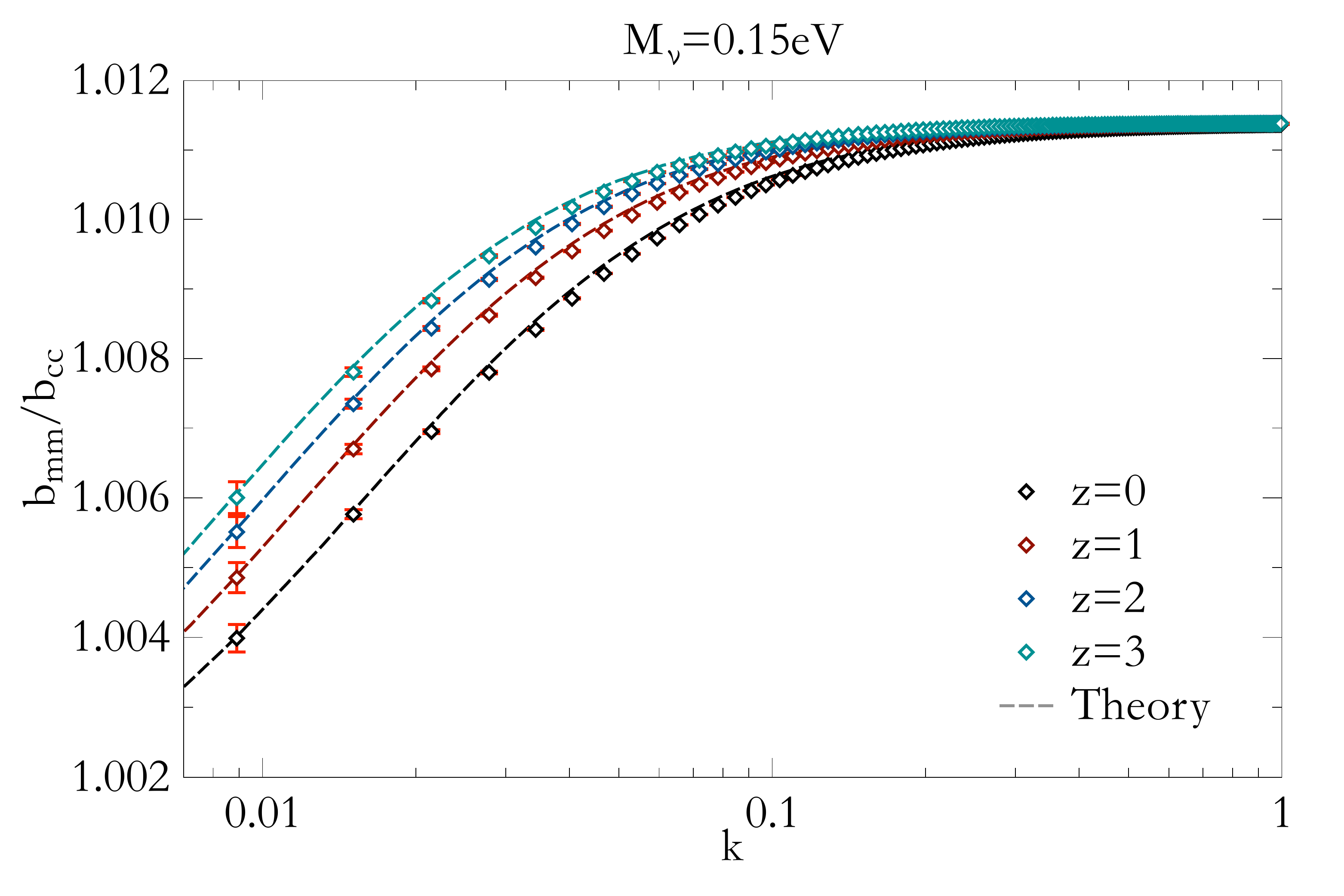}
\caption{Ratio of bias with respect to the total matter over the one of the the CDM only, $b_{\rm mm}(k)/b_{\rm cc}(k)$, from simulations for the case $M_{\nu}=0.15eV$; different lines correspond to different redshifts, with symbols corresponding to simulation outputs and dashed lines being the theoretical predictions. Errors are computed as the variance from the simulations.}  
\label{fig:bias_ratio_sim}
\end{figure}

In Figure~\ref{fig:bias_ratio_sim} we show the ratio $b_{\rm mm}(k)/b_{\rm cc}(k)$ from simulations for the model with $M_\nu=0.15$ eV neutrinos together with the linear theory prediction. The errors shown are the variance of the simulations, and they are not affected by cosmic variance as we are sampling exactly the same underlying field.

\begin{figure}
\includegraphics[width=\columnwidth]{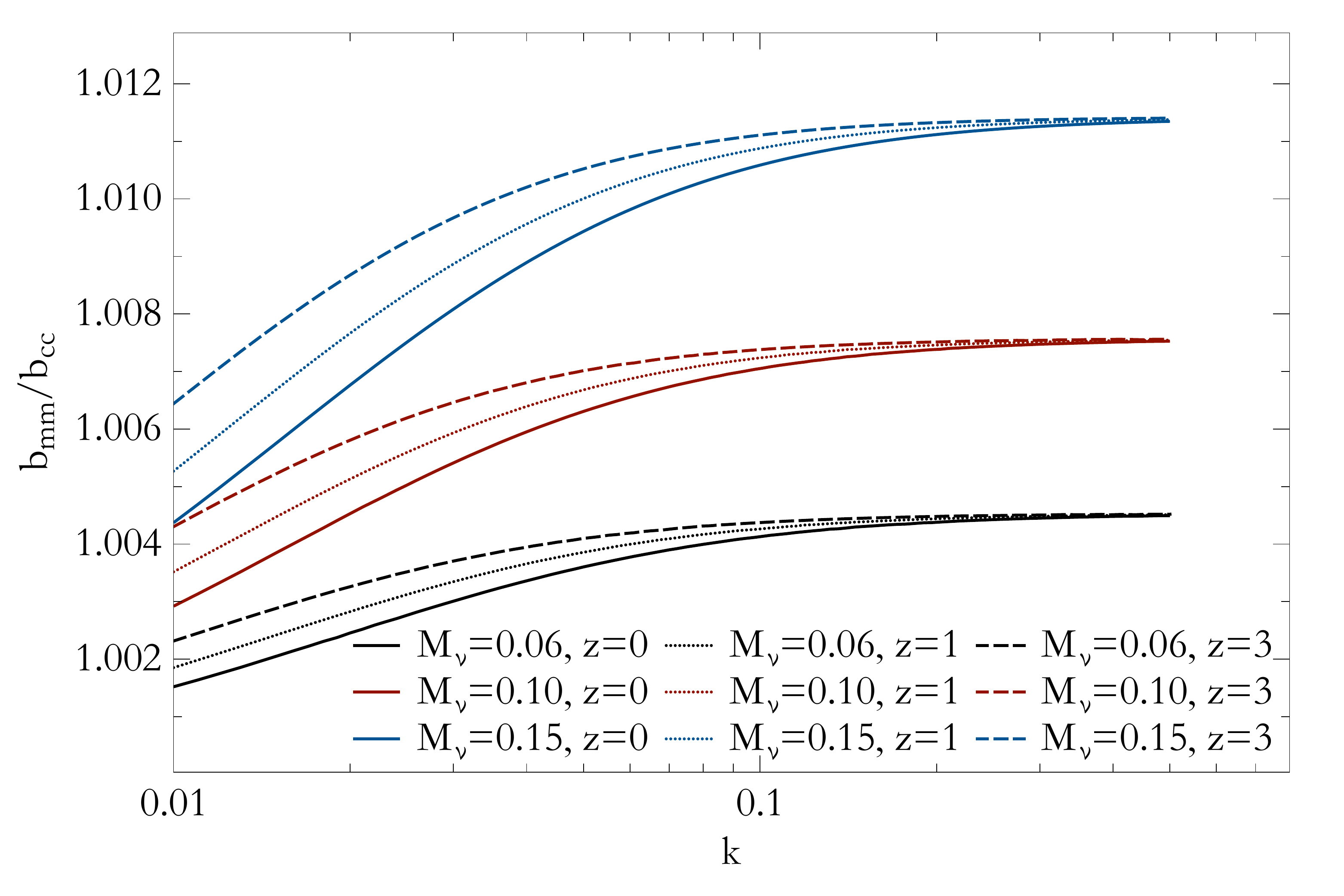}
\caption{Ratios of the bias  with respect to the of the total matter to  the one  with respect to  CDM only, for different redshifts and neutrino masses.}  
\label{fig:bias_ratios}
\end{figure}

In general the linear prediction of $b_{\rm mm}$ at every redshift and for every value of $M_{\nu}$ (using the appropriate growth factor, which is implicit here) is given by:
\begin{equation}
\frac{b_{\rm mm}(k)}{b_{\rm cc}(k)}=\frac{T_{\rm cb}(k)}{T_{\rm m}(k)} \, .
\end{equation}
 In our application, since $b_{\rm cc}$ is computed for $M_{\nu}=0$, the $b_{\rm mm}$ definition becomes:
 \begin{equation}
\frac{b_{\rm mm}(k,M_{\nu})}{b_{\rm cc}(k,M_{\nu}=0)}=\frac{T_{\rm cb}(k)}{T_{\rm m}(k)} \frac{b_{\rm cc}^{LS}(M_{\nu}=0)}{b_{\rm cc}^{LS}(M_{\nu}\neq0)}\, ,
\label{eq:5}
\end{equation}
where $b_{\rm cc}^{LS}$ is the large-scale amplitude of the bias with respect to the cold dark matter.
We find that linear theory is capable of reproducing the bias ratio from simulations down to very small, fully non-linear, scales. We then use this result to  model the ratio $b_{\rm mm}(k)/b_{\rm cc}(k)$ for other models with massive neutrinos. In Figure~\ref{fig:bias_ratios} we show that ratio for three different massive neutrino cosmologies at $z=0$, $z=1$ and $z=3$.

In what follows unless otherwise stated we use the prediction for the power spectrum of the mass and therefore the $b_{\rm mm}$  modelled as illustrated here.

\subsection{Modeling the signal using the  simulations input}
\label{sec:Pk}
The observable we consider is the halo power spectrum in Fourier (redshift) space, $P_h^s(k)$,
that we model as (see e.g.~\cite{Seo:2003}):
\begin{align}
\label{eq:pkobs}
P^s_{\rm h}(k,\mu,z,M_\nu) =  \mathcal{S}(k,\mu,z,M_\nu) \times {\rm FoG} \times P_{\delta}^r(k,z) \, ,
\end{align}
where the superscripts $r$ and $s$ indicate real and redshift-space, and the subscripts $\delta$ and $h$ stand for density and halos respectively.

The first term describes the Redshift-Space Distortions (RSD) corrections, arising from the fact that the real-space position of a source in the radial direction is modified by peculiar velocities due to local overdensities; this effect can be described at linear order as~\cite{Kaiser:1987, Hamilton:1997}:
\begin{equation}
\label{eq:kaiser}
\mathcal{S}(k,\mu,z,M_\nu) = b(k,z,M_\nu)^2 \left[1+\beta(k,z,M_\nu) \mu^2 \right]^2  \, ,
\end{equation}
where $\beta = f/b$, $b$ is the bias as described in Section~\ref{sec:bias}, and the parameter $f$ is defined as the logarithmic derivative of the linear growth factor $D$:
\begin{equation}
f = \frac{d \, \ln \, D}{d \, \ln \, a} \, .
\end{equation}
A real data analysis will need to take into account a variety of additional corrections, including geometry, general relativistic and lensing effects (see e.g.~\cite{Szalay:1997, Matsubara:1999, Papai:2008, Raccanelli:2010wa, Yoo:2010, Jeong:2011GR, Bertacca:2012, Raccanelli:2013, Raccanelli:radial}); however all these corrections will not appreciably modify our conclusions, hence we will not include a detailed modeling of them.

Given that in this work we investigate parameters that affect small scales, it is appropriate to carefully model the moderately large $k$ regime. We then include in our modeling of the power spectrum the so-called Fingers of God (FoG~\cite{Jackson:1972}), which we write as:
\begin{equation}
{\rm FoG} (k,\mu) = e^{ -\frac{k^2 \mu^2 \sigma_v^2}{H_{\rm 0}^2} } \, , \,\,\,\,\, 
\sigma_v^2 = \frac{f^2 H_{\rm 0}^2}{6\pi^2} \int P_{\theta\theta}(k) \, dk \, ;
\end{equation}
where $\sigma_v$ is the velocity dispersion and $P_{\theta\theta}(k)$ is the velocity power spectrum.

Our model is then rescaled by the factor $\frac{\left[D_{\rm A}^{\rm ref}(z)\right]^2}{\left[D_{\rm A}(z)\right]^2} \frac{H(z)}{H^{\rm ref}(z)}$, where $D_A$  is the angular diameter distance and $H$ the Hubble parameter. The superscript $^{\rm ref}$ indicates the values of the angular diameter distance and the Hubble parameter in the reference ($\Lambda$CDM) case used to compute distances from the observed angles and redshifts.

The matter power spectrum $P_{\delta}^r(k,z)$ contains the information about the primordial power spectrum.
Inflation predicts a spectrum of primordial curvature and density perturbations  that are the  seeds of cosmic structure which forms as the universe expands and cools; it's spectral index, $n_s$, is a fundamental observable for inflationary models, and it is defined as:
\begin{equation}
n_s - 1 \equiv \frac{d \, {\rm ln} \, P}{d \, {\rm ln} \, k} \, .
\end{equation}
In some inflationary models $n_s$ can be scale-dependent, and this is usually expressed as:
\begin{equation}
\alpha_s = \frac{d \, n_s}{d \, {\rm ln} \, k} \, .
\end{equation}
The quantity $\alpha_s$ is called the ``running'' of the spectral index; following standard notation and conventions (e.g.~\cite{Planck_2015}, we choose the pivot to be  $k_*=0.05/$Mpc; for a recent review of constraints on the power spectrum spectral index and its running(s), see~\cite{Munoz:2016ns}.

\subsection{Forecasting  the systematic shifts}
The Fisher matrix approach ~\cite{Fisher:1935, Tegmark:1997} is a very powerful way to  estimate or forecast uncertainties on  a model's parameters for a given experimental set up  before actually  getting the data and for this reason it has become a workhorse  tool of modern cosmology  and of survey planning. 
In this approach  one assumes that a Taylor expansion   of the likelihood ($\mathcal{L}$) around its maximum  to include second-order terms,  is a sufficiently  good approximation to estimate errors on the parameters. 
The Fisher matrix is 
\begin{equation}
F_{ij}\equiv -\left \langle\frac{\partial^2 \ln \mathcal{L} }{\partial \vartheta_{i}\vartheta_{j}} \right\rangle
\end{equation}
where $\vartheta_{i}$ denotes the model parameters.
Say that the  true underlying model (indicated by subscript T)  has  effectively $p$ extra parameters, which in the actual  model used  in the analysis (indicated by subscript U, with $n$ parameters) are instead kept fixed at incorrect fiducial values. In this case the maximum likelihood  computed  using model U will not in general be  at the correct  value for the $n$ parameters: these will  present a {\it shift} from their true values to   compensate   for the ``shift" introduced in the neglected $p$ parameters.  If the additional $p$ parameters are shifted by  $\Delta \psi_\zeta$, the Fisher matrix approach allows one to estimate the resulting systematic shifts in the other $n$ parameters by using~\cite{Tayloretal2007, HeavensKitchingVerde}:
\begin{equation}
\Delta \vartheta_i=-\sum_{j,\zeta}(F_U^{-1})_{ij}G_{j \zeta}\Delta \psi_{\zeta}
\end{equation}
where $F_U$ denotes the  $n\times n$ Fisher matrix of model U, and $G$ is a  subset of the  $(n+p)\times(n+p)$ Fisher matrix $F_T$.   For our applications, the extra parameters are those that describe the scale dependence in the halo bias introduced by non-zero neutrino masses. In other words, in the model T, $M_{\nu}>0$, while for model U, neutrinos have masses but in the halo bias expression $M_{\nu}$ is set to 0.

We write the Fisher matrix for the power spectrum as
\begin{align}
\label{eq:FM}
F_{ij}=\int \mathrm dz \, \mathrm dk \, \mathrm d\mu \, k^2 \frac{V_{\rm eff}}{8\pi^2} \frac{\partial \ln P}{\partial \vartheta_i}\frac{\partial \ln P}{\partial \vartheta_j} \, ,
\end{align}
where $\vartheta_{i}$ runs over the parameters of interest $\{n_s,\,\alpha_s,\, M_\nu, b, \Omega_m\}$, and the effective volume of the survey in the $z$-th redshift bin is:
\begin{equation}
V_{\rm eff}(k,\mu,z,M_\nu) = V_s\left[\frac{\bar{n}_g(z) P(k,\mu,z,M_\nu)}{1+\bar{n}_g(z) P(k,\mu,z,M_\nu)}\right]^2 \, ;
\end{equation}
$V_s$ is the volume of the survey and $\bar{n}_g$ is the mean comoving number density of galaxies.

\begin{figure}
	\centering
	\includegraphics[width=0.95\columnwidth]{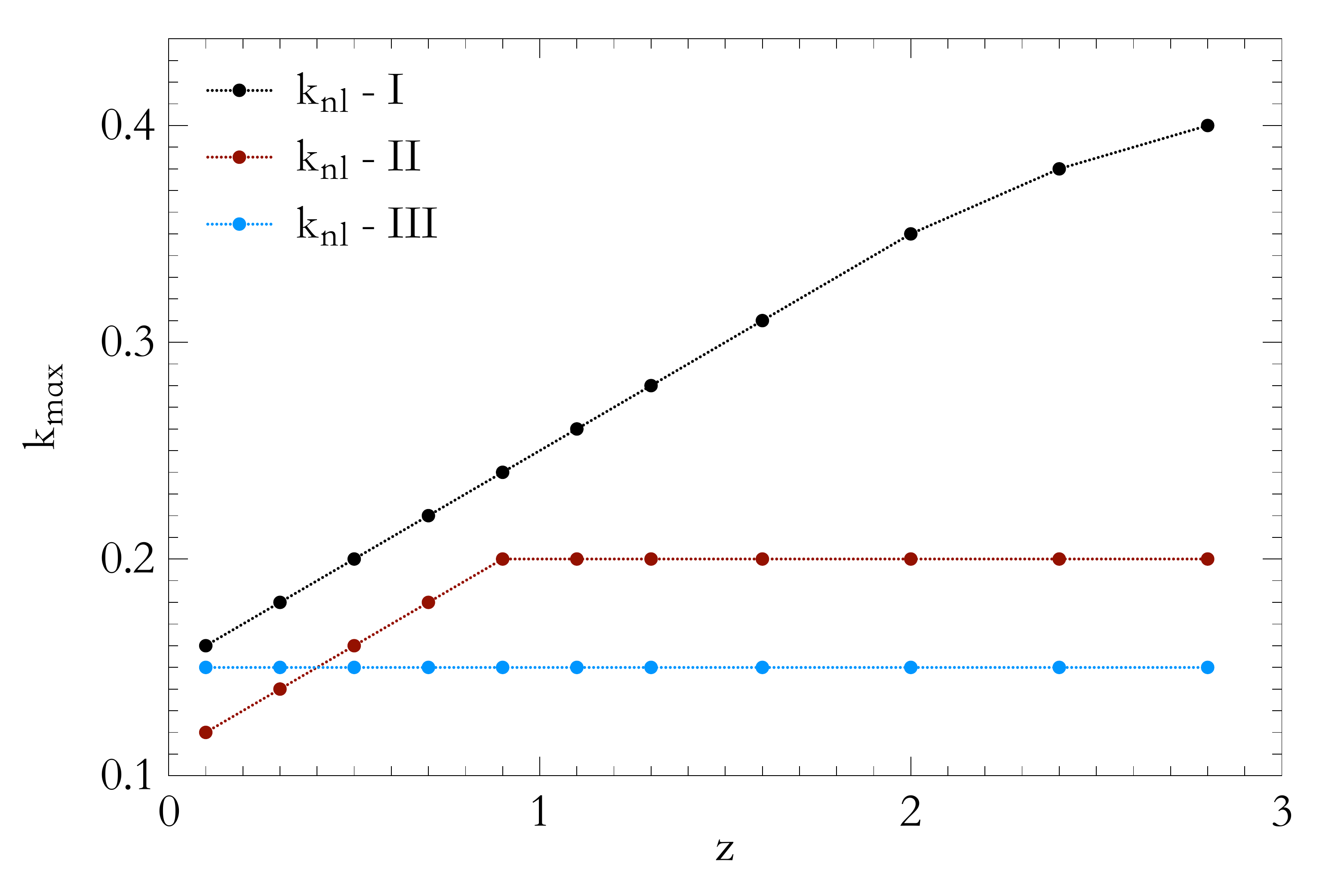}
	\caption{The three models for the minimum scale used for computing our results.}
	\label{fig:knl}
\end{figure}

Equation~\eqref{eq:FM} involves an integral over the wavenumber $k$; the largest scale that can be probed is determined by the geometry of the survey, $k_{\rm min} = 2\pi V_s^{-1/3}$; for the smallest scale $k_{\rm max}$, we consider three cases, to illustrate the effect of neutrino bias for different assumptions and non-linear modeling that could be chosen for future analyses, which we name $k_{\rm nl}-I, II, III$. In Figure~\ref{fig:knl} we show the redshift dependence of $k_{\rm max}$ in these three cases. Note that in case I  the maximum $k$ increases in redshift so that the r.m.s of the density fluctuations is constant in redshift and has the same value as the one for $k_{\rm max}=0.16$ $h$/Mpc at $z=0$.
Case II is more conservative, having $k_{\rm max}=0.12$ $h$/Mpc at $z=0$; $k_{\rm max}$ initially grows in redshift to keep $\Delta^2(k_{\rm max})$ constant but then it saturates at $k_{\rm max}=0.2$.
Case III is some more simplistic and very conservative assumption, as it keeps $k_{\rm max}=0.15$ $h$/Mpc, constant in redshift. 

For practical purposes, the difference between $F_T$ and $F_U$ is simply given by a different model for the galaxy bias.

The systematic change in $P(k)$ will change the log likelihood as:
\begin{equation}
{\rm ln} \, \mathcal{L}^U = {\rm ln} \, \mathcal{L}^T - \frac{1}{2} \sum_k \frac{\Delta P}{P};
\end{equation}
if we assume that the change is small, we can Taylor expand around the true parameter values, which gives:
\begin{equation}
\partial_j {\rm ln} \, \mathcal{L}(\vartheta_i^T) = \partial_j {\rm ln} \, \mathcal{L}(\vartheta_i^U) + \partial_i \partial_j {\rm ln} \, \mathcal{L}(\vartheta_i^U)\Delta\vartheta_i \, ,
\end{equation}
where $\Delta\vartheta_i$ is the change in the $\vartheta_i$ parameter from the T to the U models. Therefore, the change in the parameters is given by:
\begin{equation}
\Delta\vartheta_i = - \left[\partial_i \partial_j {\rm ln} \, \mathcal{L}^U(\vartheta_i^U)\right]^{-1} \left[ \frac{1}{2} \sum_k \frac{\partial P}{\vartheta_j}\frac{\Delta P}{P^2} \right] \, ,
\end{equation}

so that, using the Fisher formalism of Equation~\eqref{eq:FM}, we can compute how the best-fit for the parameters is modified by using the scale-dependence of the bias in the case of non-zero neutrino mass using:
\begin{equation}
\Delta(\vartheta_i) = \sum_j (F^{-1})_{ij} \delta_j(\vartheta) \, ,
\end{equation}
where $(F^{-1})_{ij}$ is the $\{ij\}-th$ element of the Fisher matrix of Equation~\ref{eq:FM} and:
\begin{equation}
\delta_j(\vartheta_i) = \nsum_i \int \frac{\Delta P}{P^2} \frac{\partial P}{\partial \vartheta_j} \frac{k^2 \, V_{\rm eff}}{\, 8\pi^2} \, ,
\end{equation}
where $\Delta P$ is the difference between the T and U power spectra.

The magnitude of the systematic shift, although important, does not have a straightforward interpretation without an estimate of the  corresponding statistical error.
This approach yields both the systematic shift and the statistical errors. The statistical errors however depend on  which  and how many parameters  are being considered and varied in the analysis.  For example  when analysing real survey data  within the so-called full power spectrum shape approach, one might assume that shape (i.e. space-dependence) and amplitude of the bias is unknown in every redshift bin and marginalise over e.g., the coefficients of second or third order polynomials for the bias as a function of $k$ independently in every redshift bin. This would be a very conservative approach yielding relatively large statistical errors on cosmological parameters. At the other end of the spectrum one could assume that the bias shape and amplitude is perfectly known at all redshift. This would be an overly aggressive approach. In reality there will probably be priors imposed on the bias amplitude and shape and their evolution with redshift, the analysis will then marginalise over these priors. These priors are however yet unknown. 
In the following Section we will consider two different approaches and present results for a more aggressive and a more conservative choice.
To compute our results, we use a Fisher matrix with $\{n_s, \alpha_s, M_\nu, A_{\rm bias}(z), \Omega_m \}$, where $A_{\rm bias}(z)$ is the bias amplitude, over which we marginalize. In our formalism, this gives the Fisher matrix for both models T and U.
We consider two different cases for the approach we follow to estimate the statistical errors: one in which we assume that the shape and redshift evolution of the bias is given by the simulations and therefore known, but an overall bias amplitude is unknown and we marginalised over one $A_{\rm bias}$ (no redshift dependence); in the second case, we marginalize over the  bias amplitude independently in  three  broad redshifts bins thus including in our Fisher matrix $A_{\rm bias}(z_1), A_{\rm bias}(z_2), A_{\rm bias}(z_3)$.

By choosing this set of  four cosmological parameters, we implicitly assume that other cosmological parameters are kept fixed. Within a $\Lambda$CDM model these are the parameters that affect the power spectrum, with the exception of $\Omega_b$, which however is tightly constrained by CMB observations. 
Our results will be presented when including or not Planck priors; in the case with priors, we first compute the power spectrum Fisher matrix and then we sum the Planck matrix provided by  the Planck Legacy Archive, using the matrix elements for the $\{n_s, \alpha_s, \Omega_m \}$ parameters, before inverting it to obtain the forecasted constraints.

\section{Results}
\label{sec:results}
We consider a variety of forthcoming galaxy surveys such as DESI\footnote{http://desi.lbl.gov/}, the ESA satellite Euclid\footnote{http://sci.esa.int/euclid/} and the planned NASA satellite WFIRST\footnote{https://wfirst.gsfc.nasa.gov/}; we model the survey specifications according to, respectively,~\cite{desispecs, euclid, wfirst}.
In particular, for DESI we use the combination of the Bright Galaxy Survey(BGS) and the main galaxy survey, using 14,000 deg.$^2$, resulting in a redshift range $0.05<z<1.85$; when marginalizing over three values of the bias, we consider three large bins centered at $z=0.5, 1.0, 1.5$.
For Euclid, we assume the $n_3$ redshift distribution of~\cite{euclid}, with $0.9<z<1.9$ over 15,000 deg.$^2$; the marginalized bias case is calculated on bins centered at $z=1.1, 1.4, 1.7$.
As for WFIRST, we combine the number density of H$\alpha$ and O{\small III} observations in order to obtain a very deep $1.0<z<3.0$ survey over 2,000 deg.$^2$, using bins centered at $z=1.2, 1.9, 2.6$.
In this way we will cover low- and high- redshift, wide and narrow, surveys, and therefore understand if the effect of neglecting the neutrino mass-induced scale dependence of the bias is more or less important for particular parts of the observational paratemer space.

\begin{figure*}[htb]
\centering
\includegraphics[width=0.68\columnwidth]{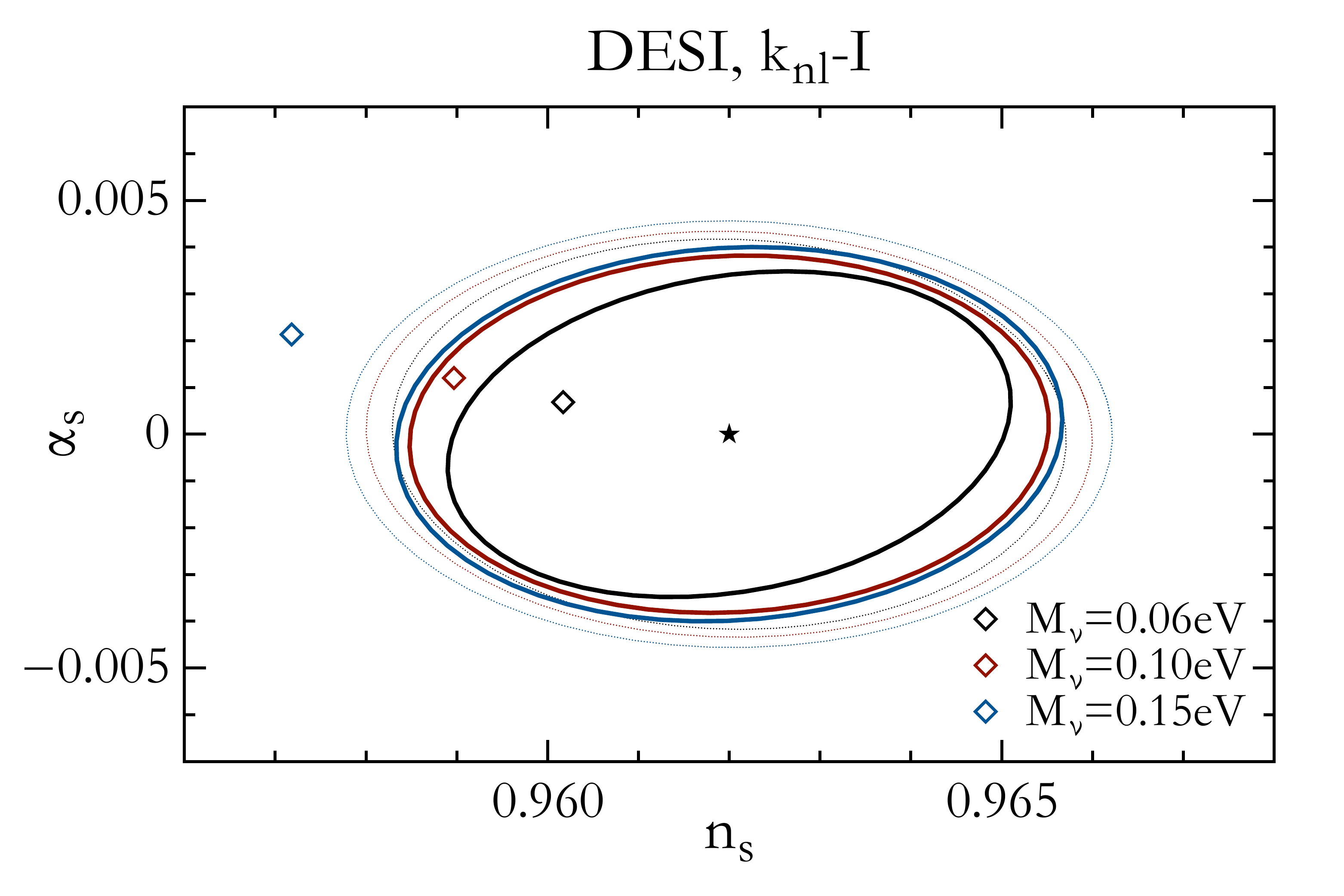}
\includegraphics[width=0.68\columnwidth]{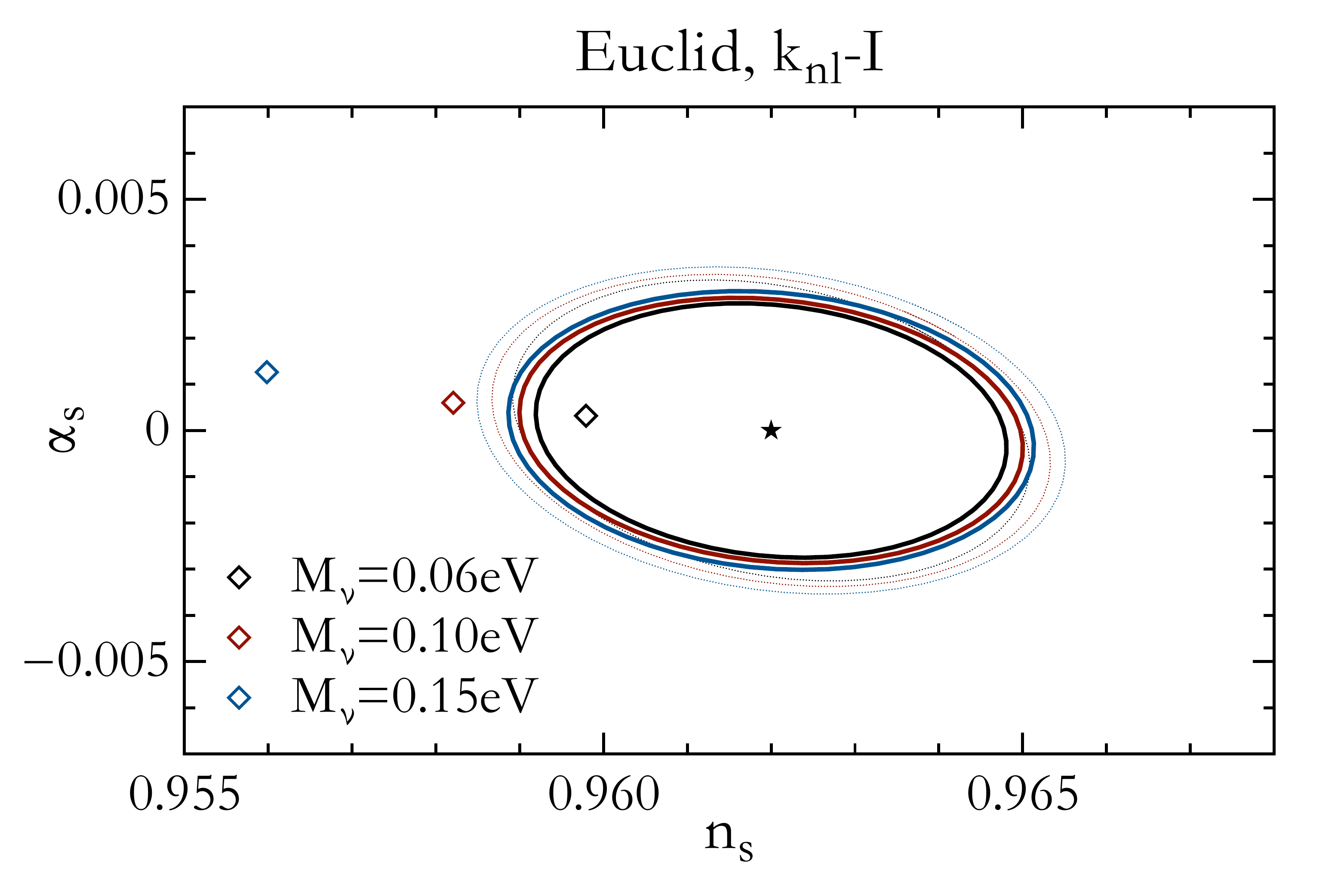}
\includegraphics[width=0.68\columnwidth]{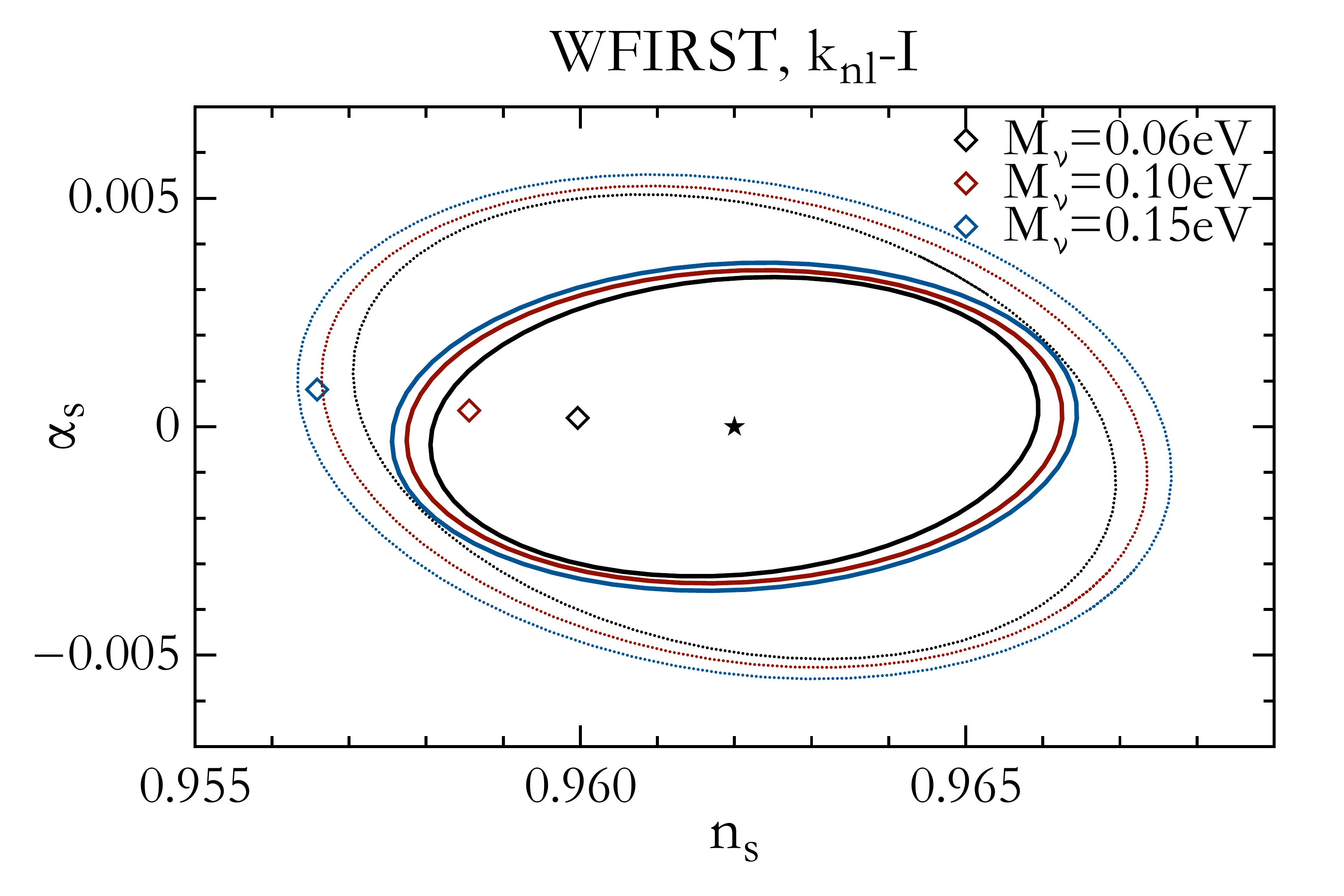}
\includegraphics[width=0.68\columnwidth]{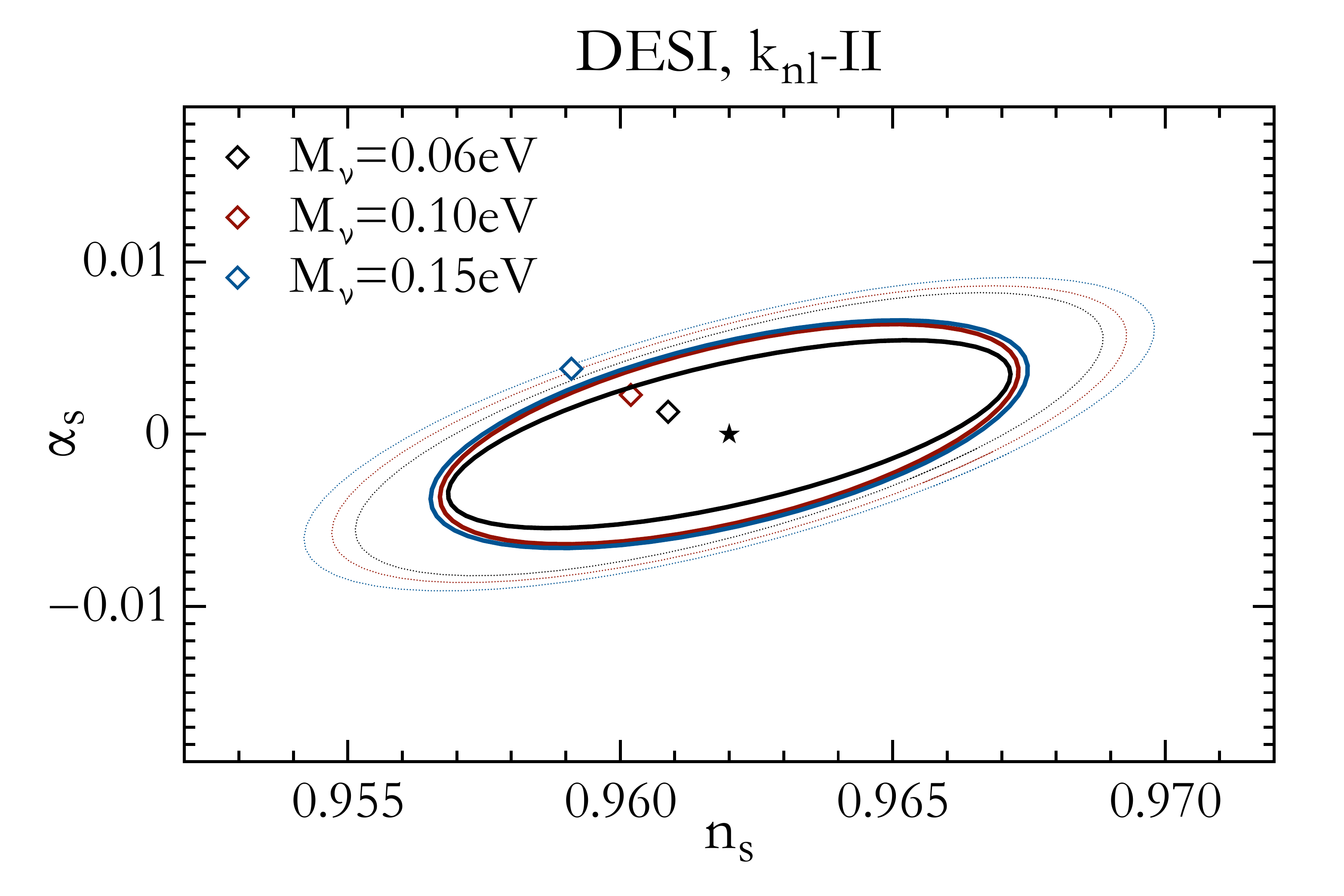}
\includegraphics[width=0.68\columnwidth]{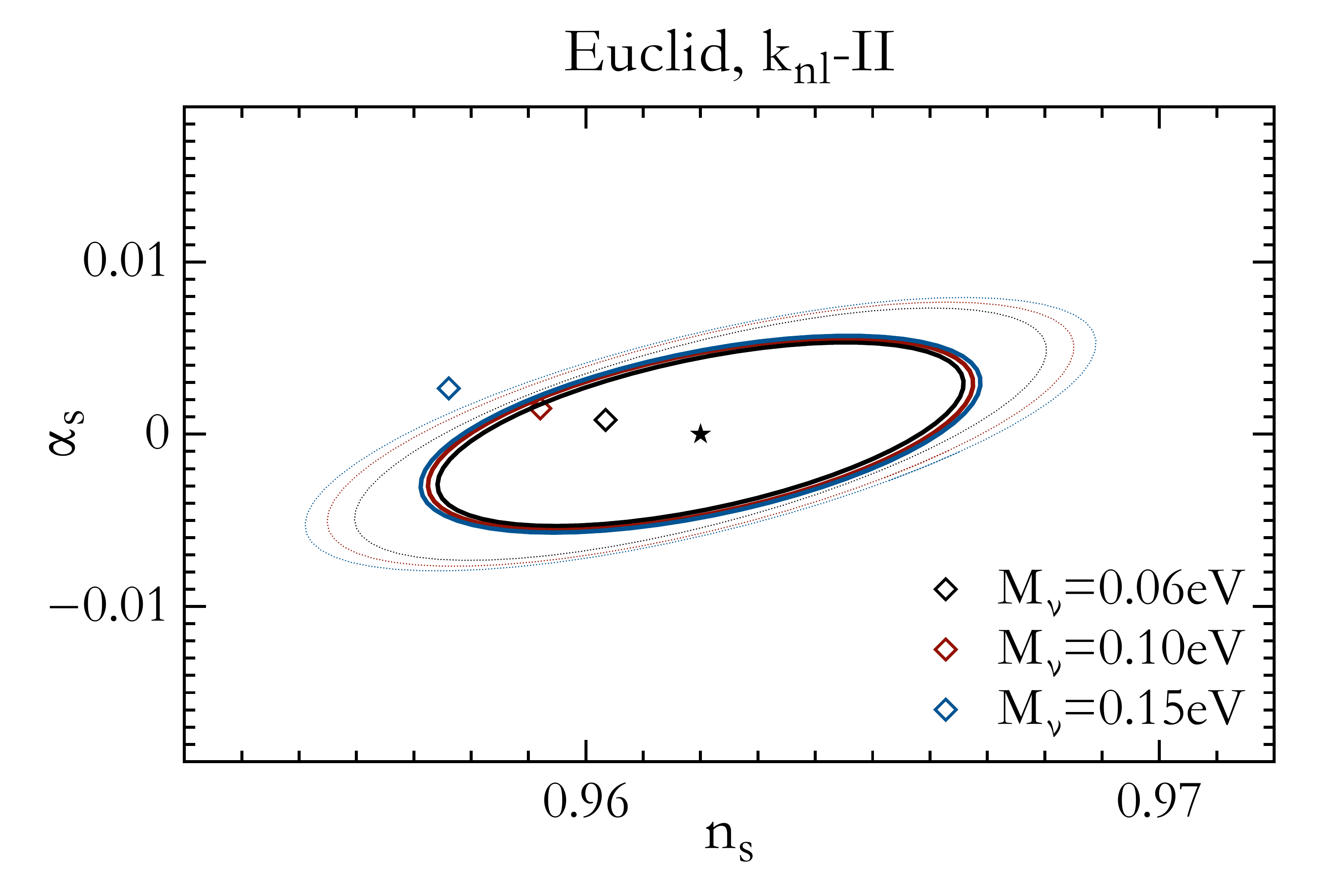}
\includegraphics[width=0.68\columnwidth]{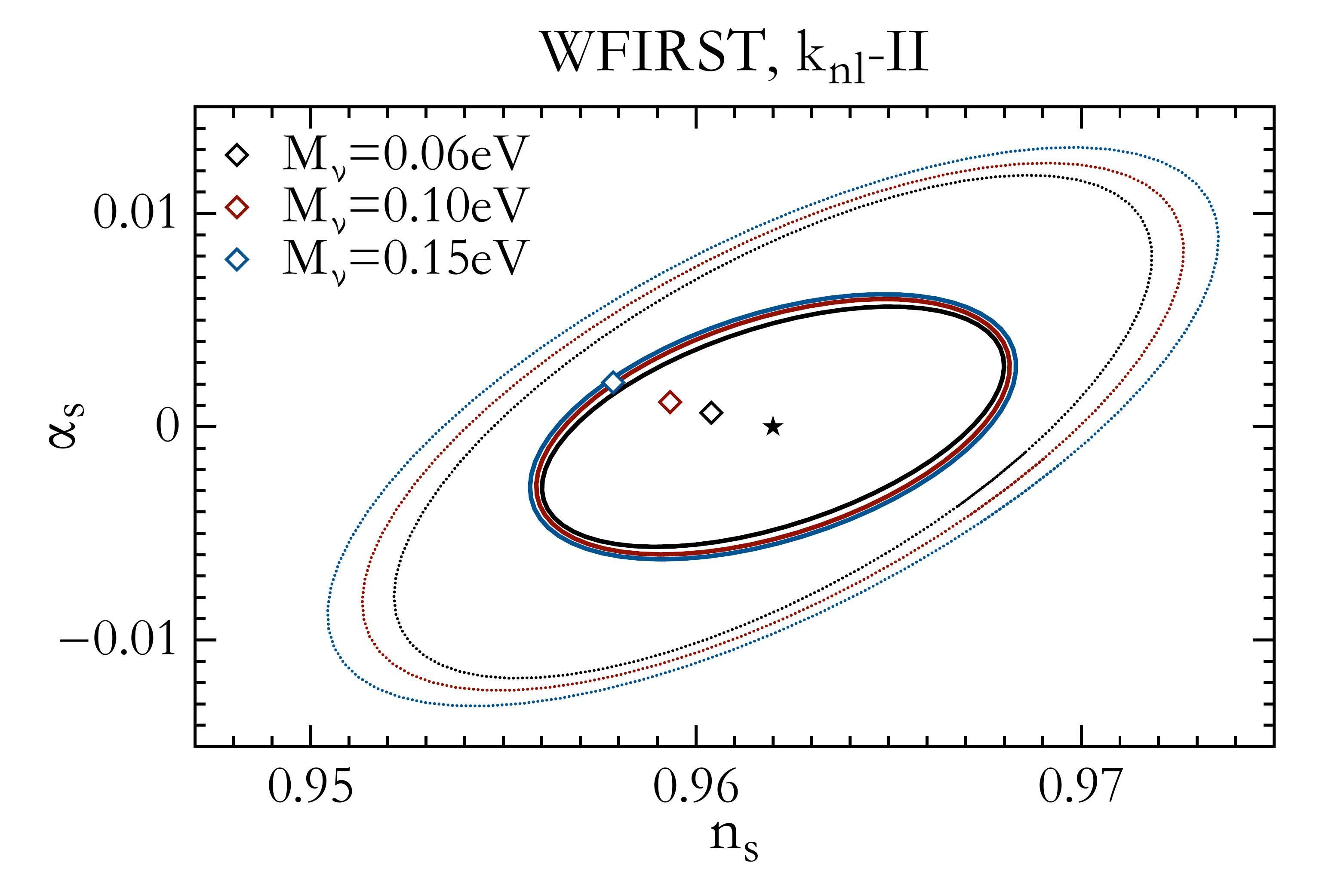}
\includegraphics[width=0.68\columnwidth]{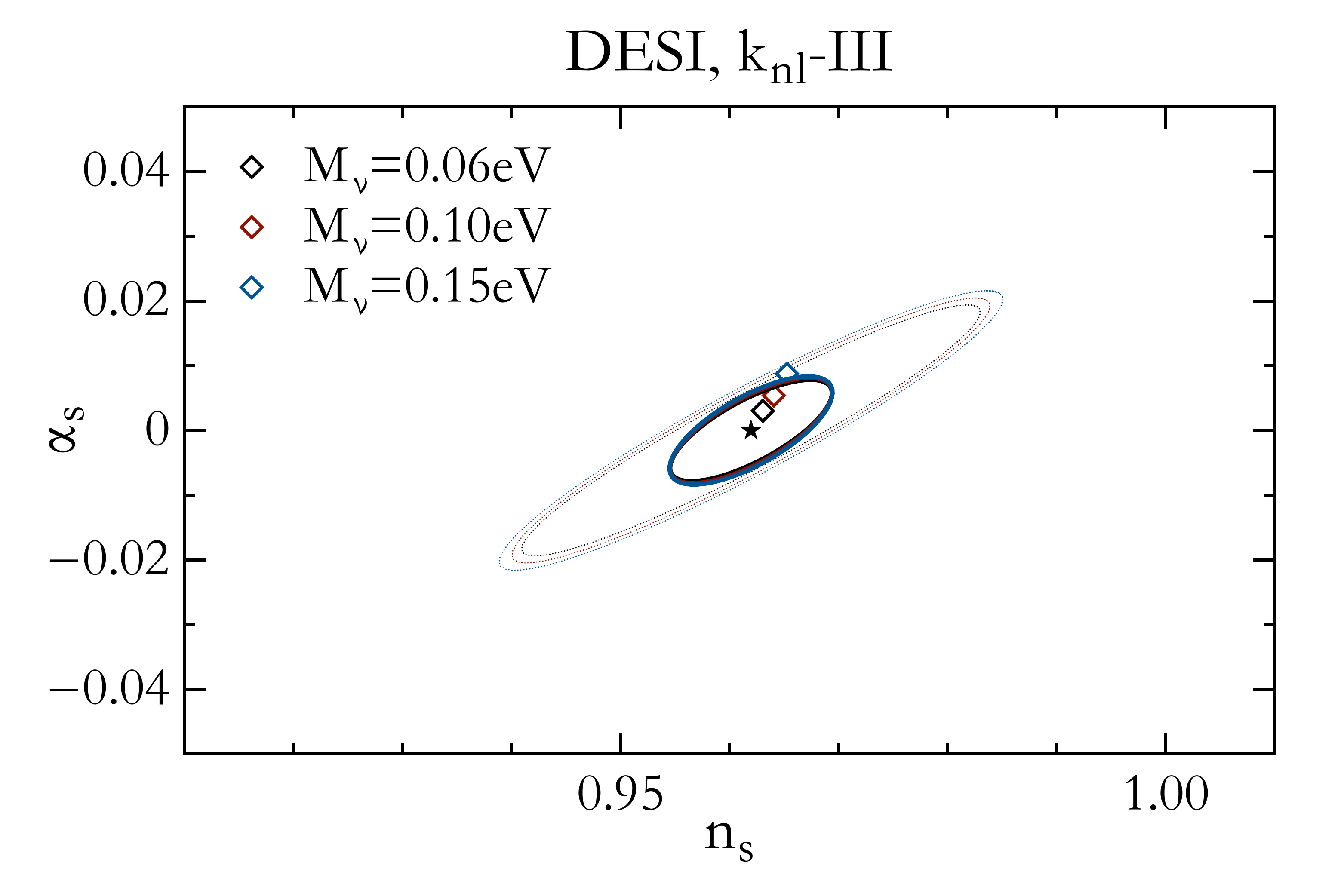}
\includegraphics[width=0.68\columnwidth]{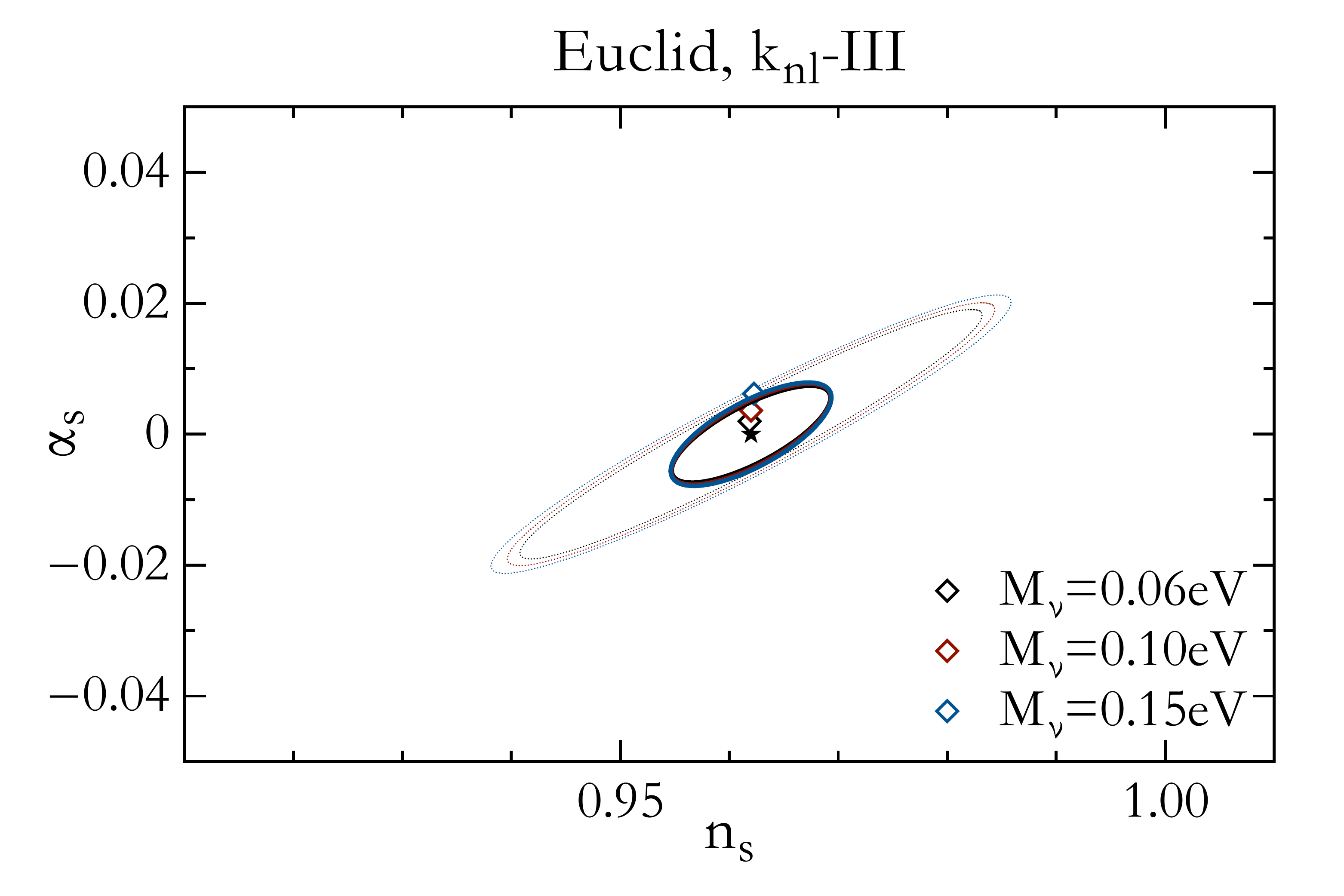}
\includegraphics[width=0.68\columnwidth]{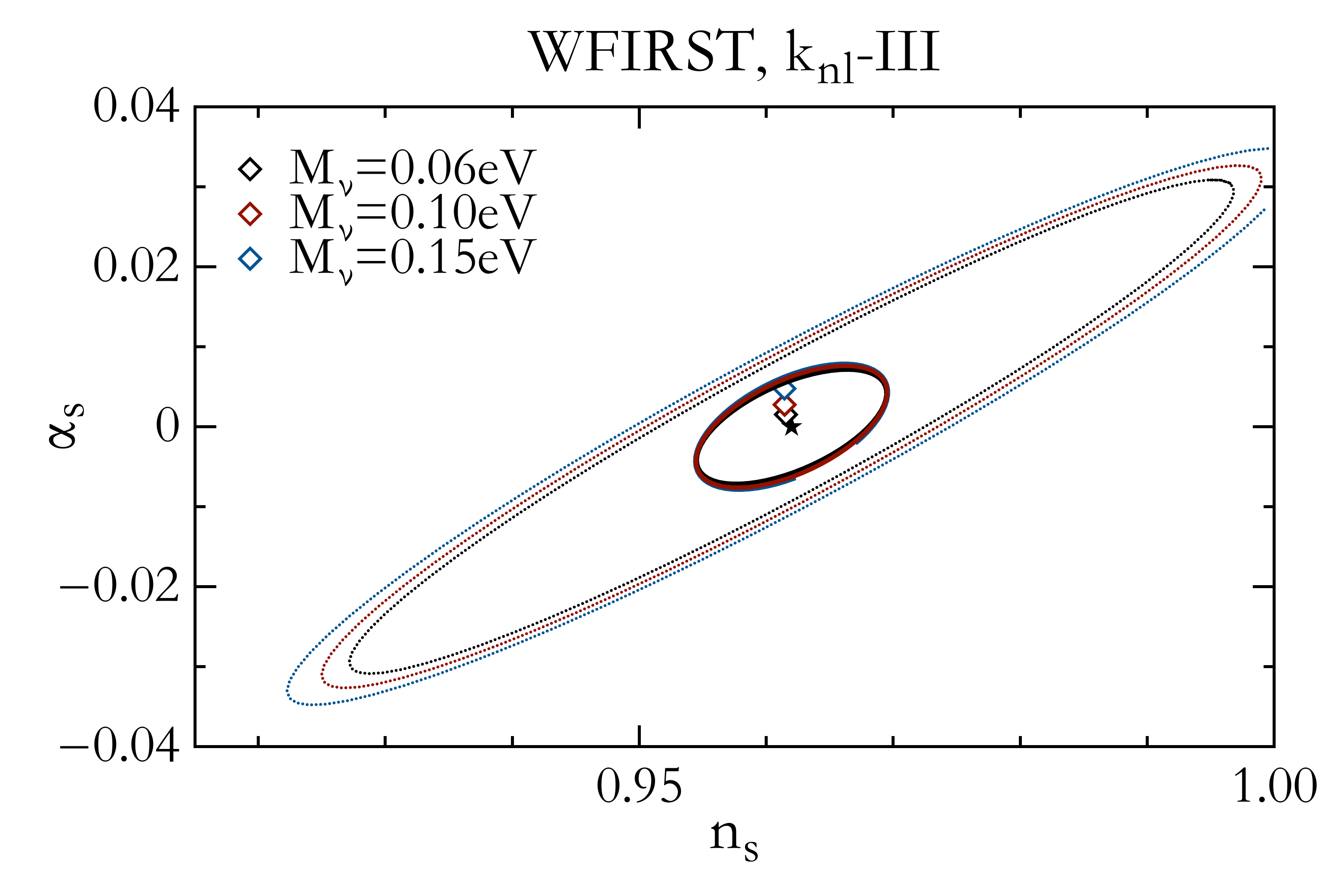}
\caption{Shift and forecasted  1-$\sigma$ joint error ellipses  in $\{n_s,\alpha_s\}$ parameters for DESI, Euclid and WFIRST-like surveys, for different values of  fiducial neutrino mass and different choices of  maximum wavenumber $k$ (see text for details). Error ellipses are entered around the fiducial model, indicated by the star symbol; solid lines show ellipses including Planck priors, thin dotted ones the constraints without Planck priors.}
\label{fig:shift_nsas}
\end{figure*}

Now we present the shift in best-fit parameters in the  two-dimensional $\{n_s,\alpha_s\}$ and $\{n_s, M_\nu\}$ planes, with the predicted error ellipse centered around the fiducial model, for the case where we marginalize over one single bias amplitude.
We show results for the three future surveys considered, for the different maximum scales as in Figure~\ref{fig:knl}, for values of the sum of neutrino masses of $M_\nu$=0.06, 0.10, 0.15eV.
In Figure~\ref{fig:shift_nsas} we plot the shift for $\{n_s,\alpha_s\}$, for the single bias amplitude marginalization case; as it can be seen, results vary considerably with different choices of the maximum wavenumber $k$ included in the analyses.

Shifts are generally larger for the wide (DESI and Euclid) surveys than for WFIRST. While the change in the best-fit for the running of the spectral index is usually small, the one for the spectral index itself if large: depending on the value of neutrino mass assumed, it can exceed 1-sigma constraints, even when not including Planck priors, for the most aggressive $k_{nl}$ we consider.

\begin{figure*}[htb]
\centering
\includegraphics[width=0.68\columnwidth]{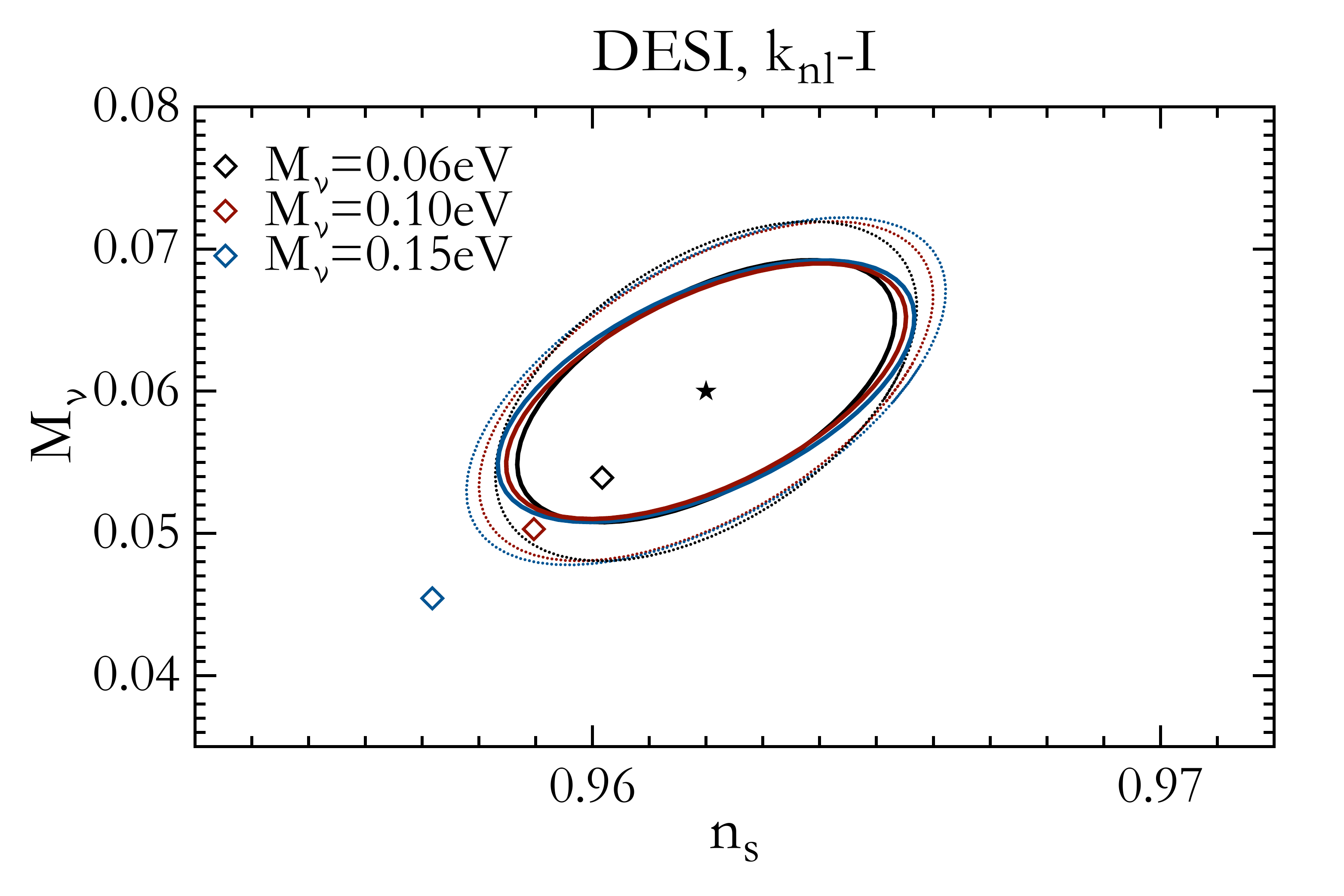}
\includegraphics[width=0.68\columnwidth]{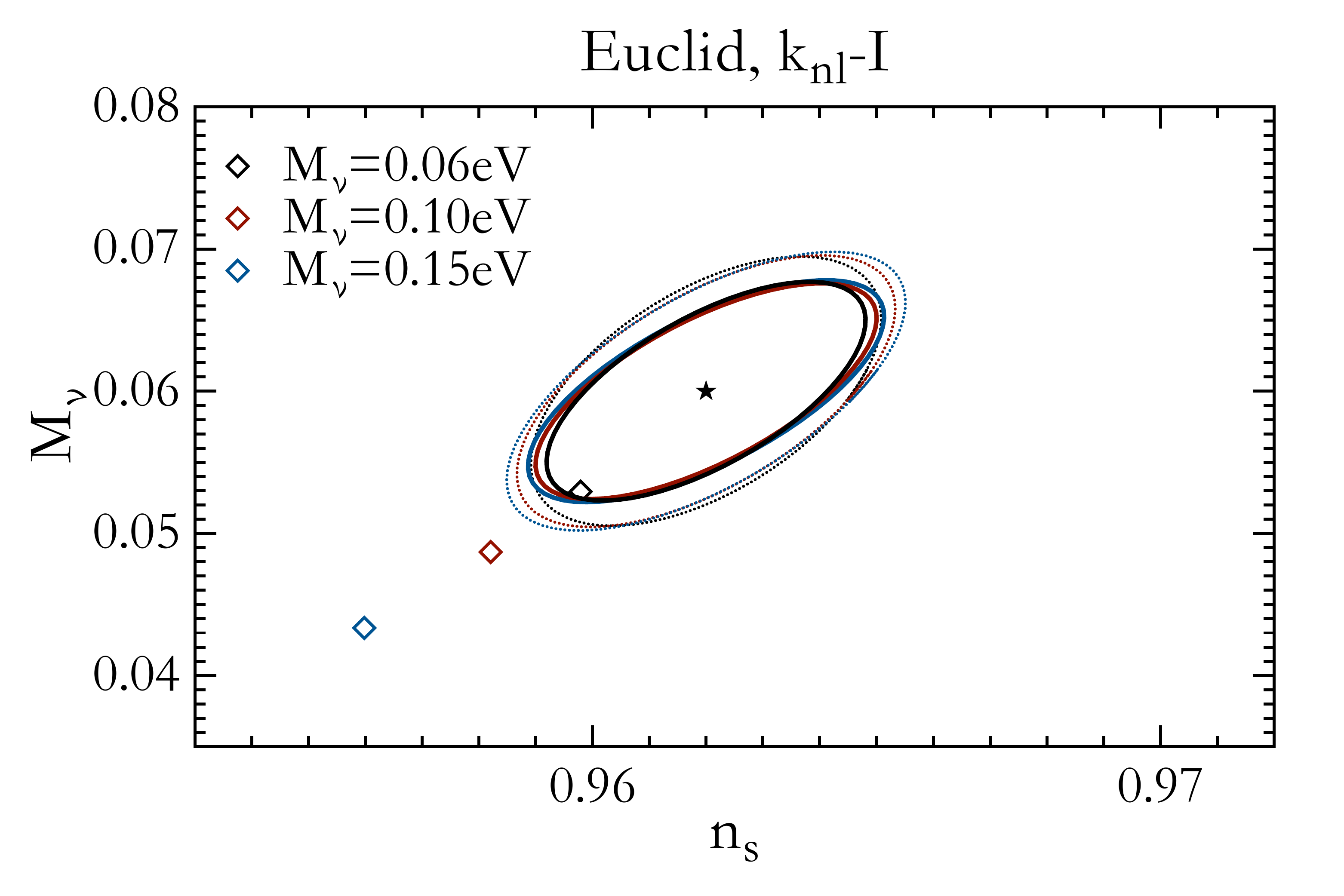}
\includegraphics[width=0.68\columnwidth]{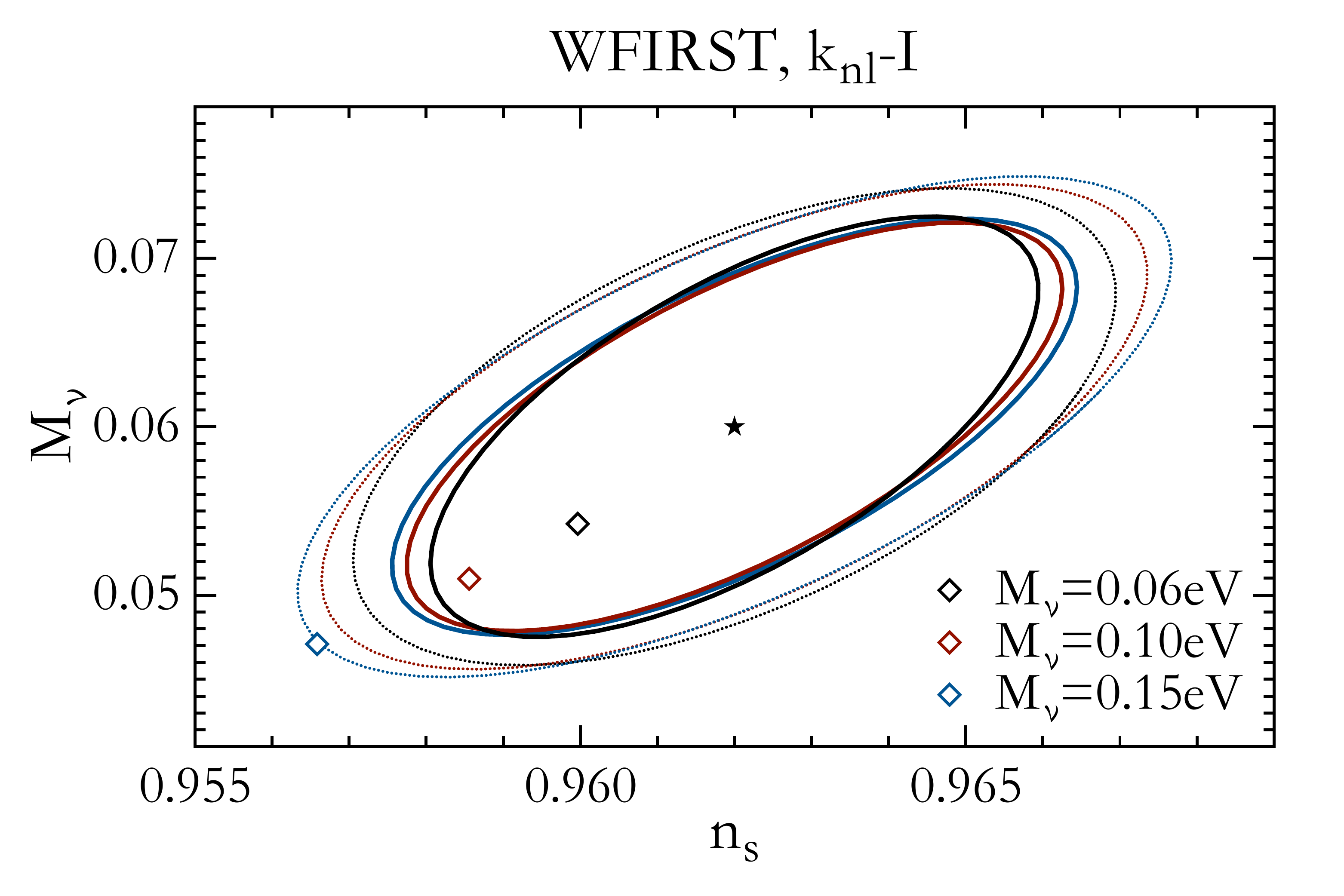}
\includegraphics[width=0.68\columnwidth]{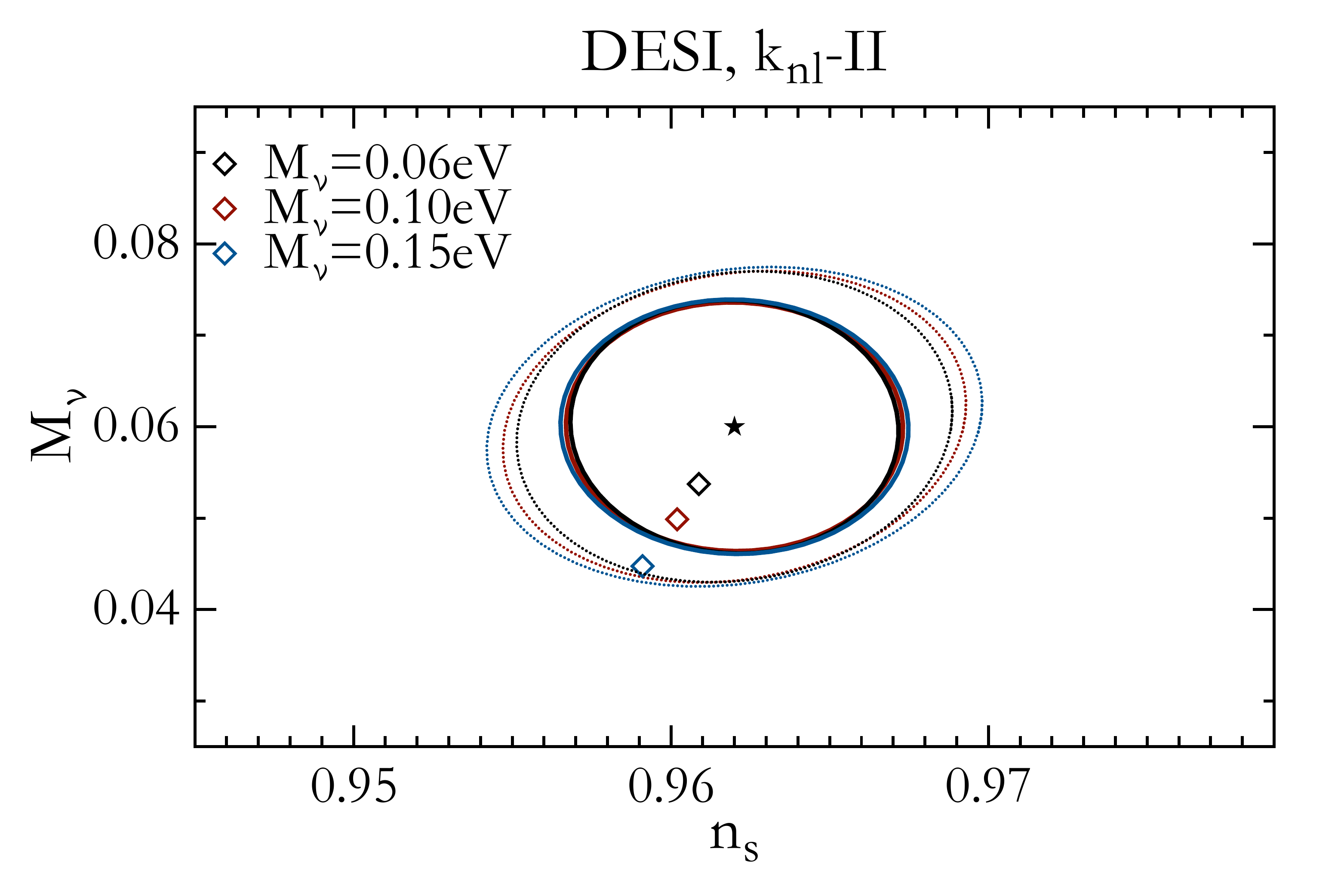}
\includegraphics[width=0.68\columnwidth]{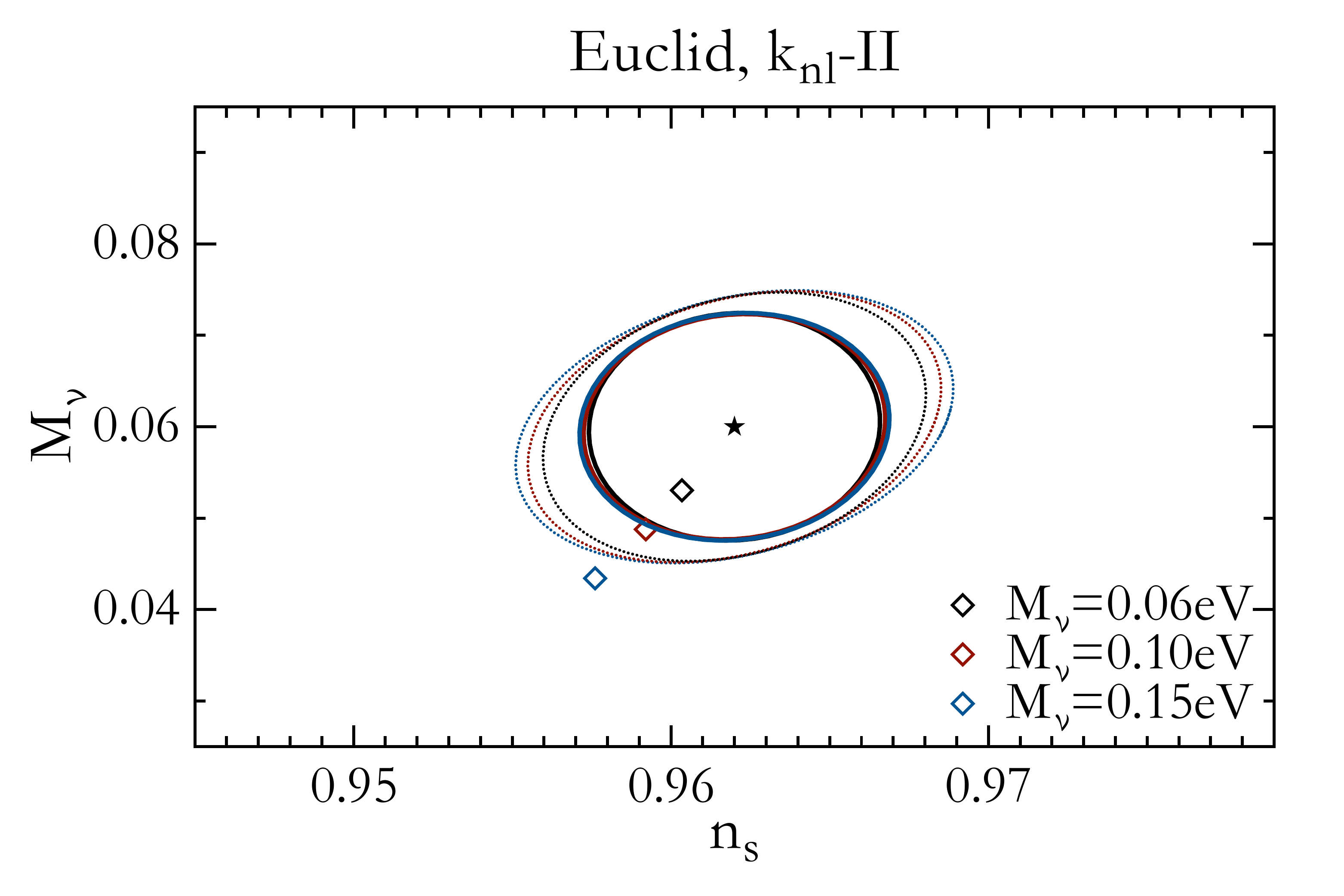}
\includegraphics[width=0.68\columnwidth]{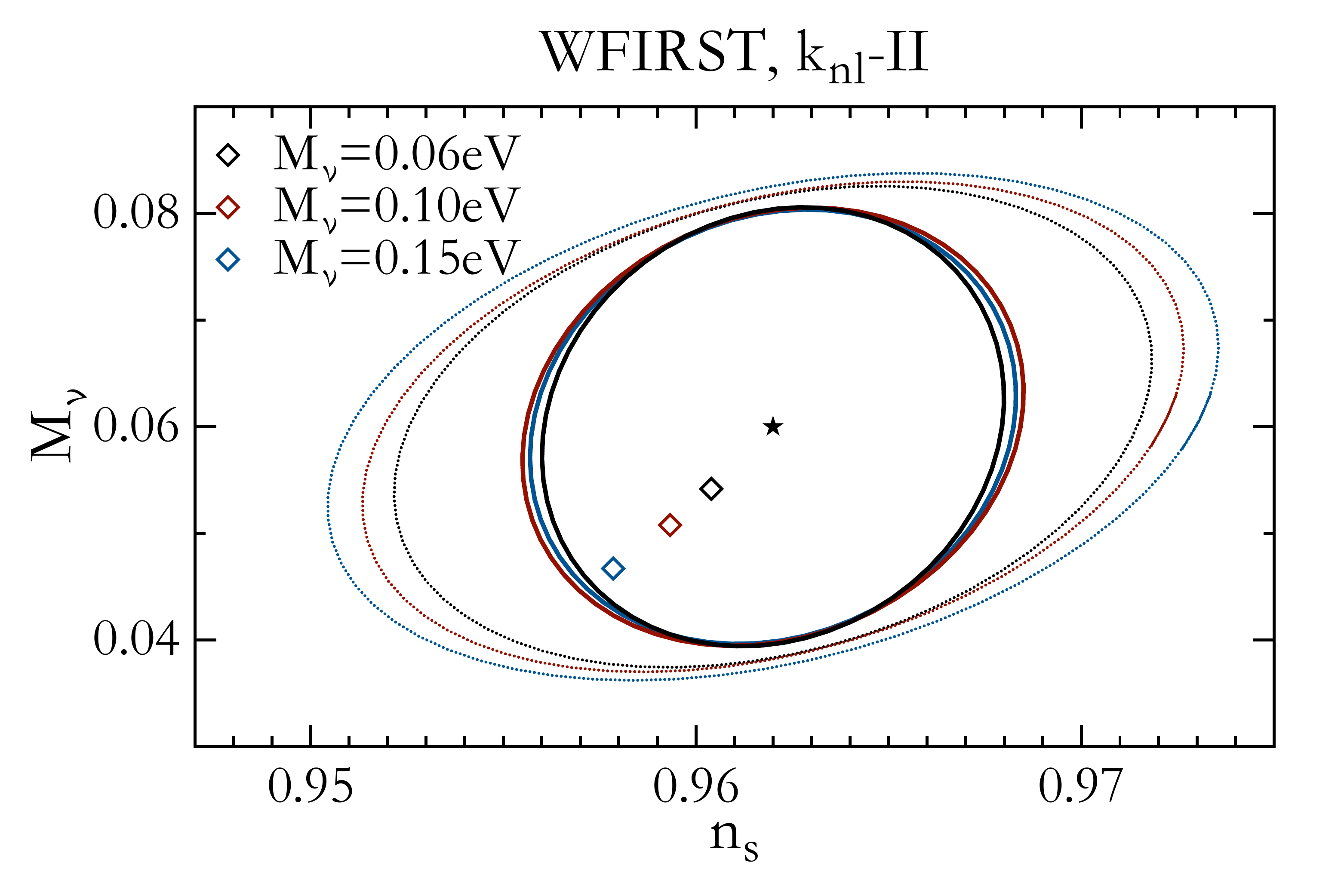}
\includegraphics[width=0.68\columnwidth]{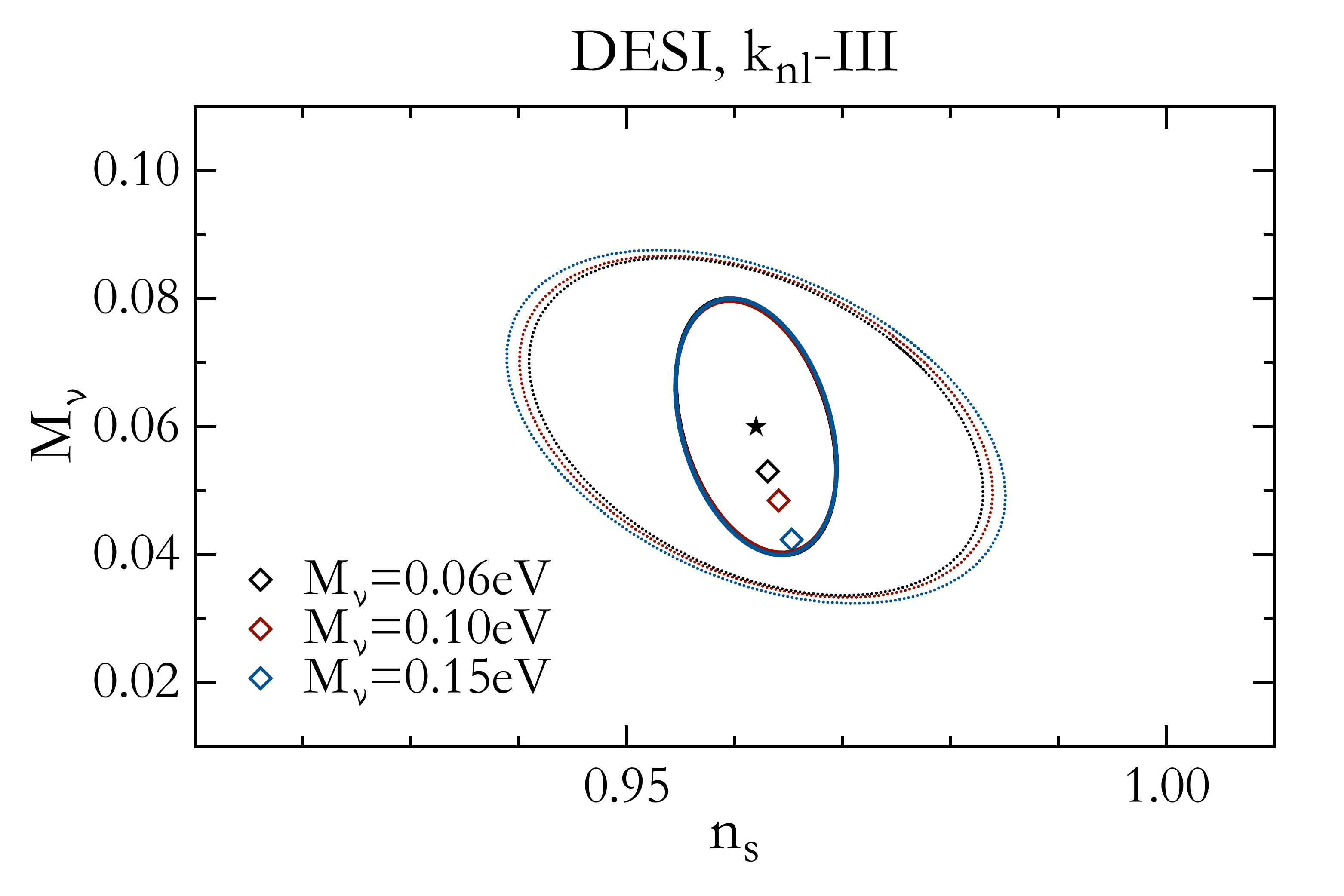}
\includegraphics[width=0.68\columnwidth]{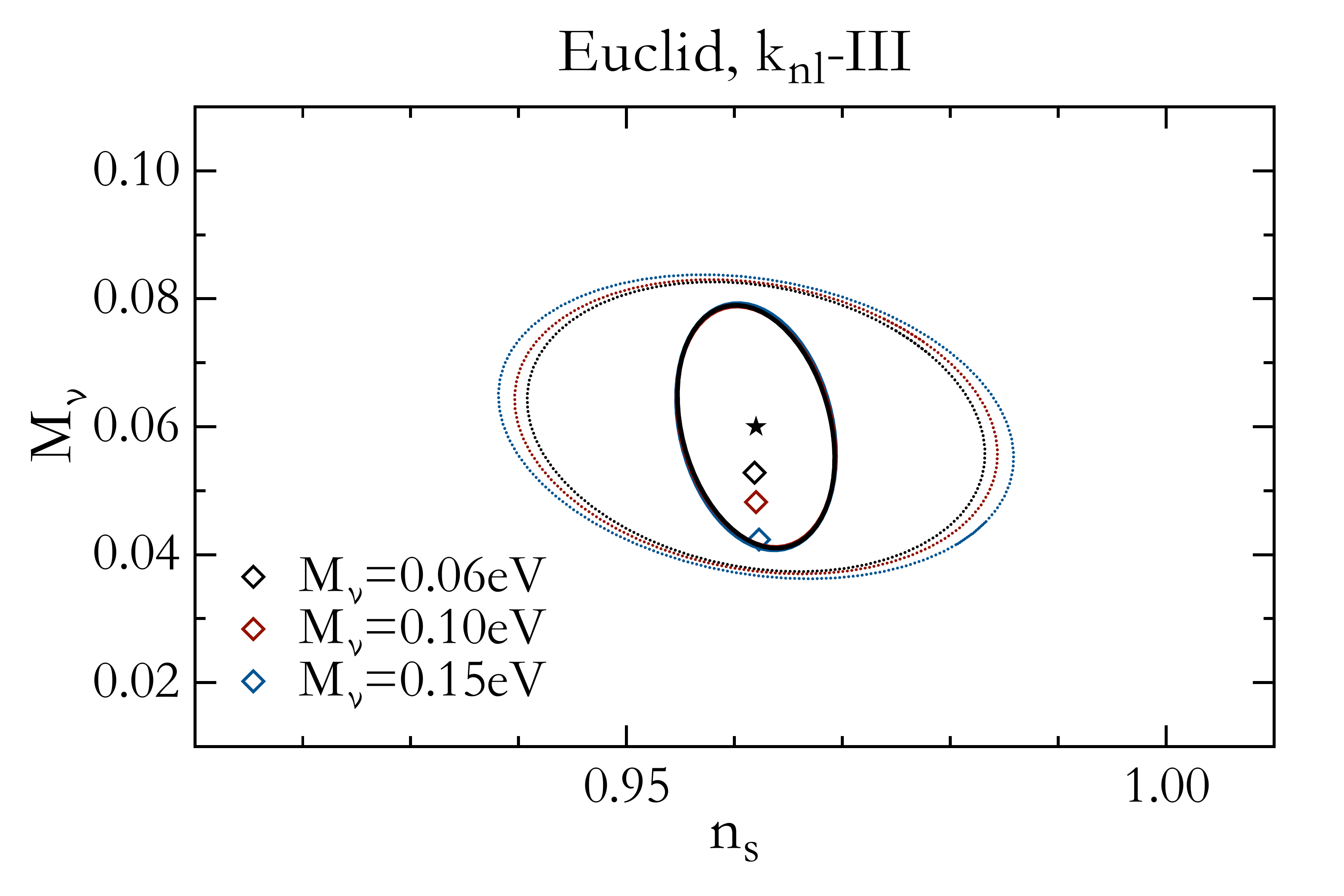}
\includegraphics[width=0.68\columnwidth]{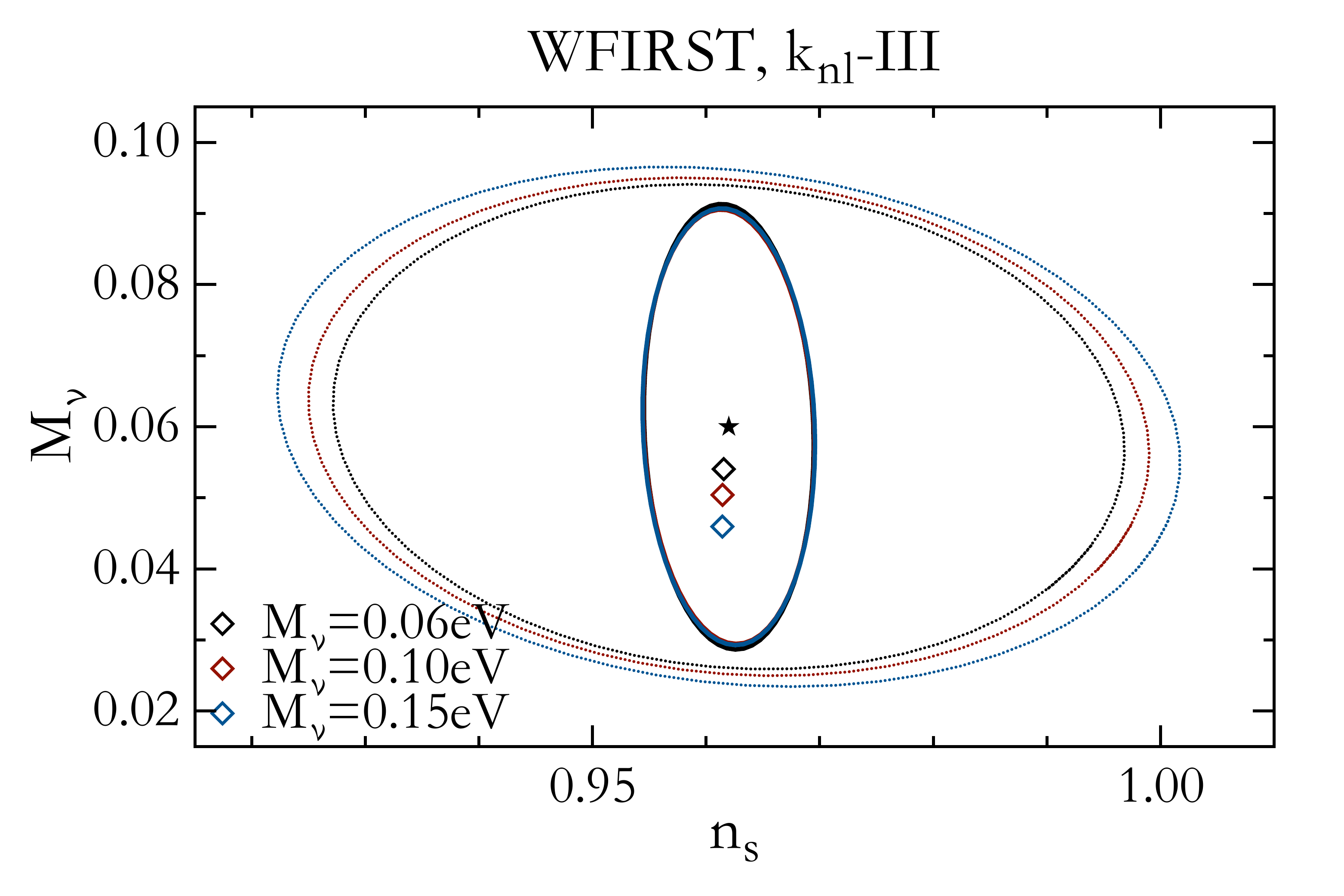}
\caption{Shift and forecasted  1-$\sigma$ joint error ellipses  in $\{n_s, M_\nu\}$ parameters for DESI, Euclid and WFIRST -like surveys, for different values of  fiducial neutrino mass and different choices of  maximum wavenumber $k$ (see text for details). Error ellipses are entered around the fiducial model, indicated by the star symbol; solid lines show ellipses including Planck priors, thin dotted ones the constraints without Planck priors.}
\label{fig:shift_nsMnu}
\end{figure*}

In Figure~\ref{fig:shift_nsMnu} we show the same results but for the combination $\{n_s, M_\nu\}$: in this case shifts are considerable not only for $n_s$, but for $M_\nu$ as well. Even for the most conservative $k_{nl}$ case, shifts are close to 1-sigma errors when including Planck priors, and are at least a non-negligible fraction of the forecasted errors for the galaxy surveys alone, indicating that the effect of neutrino mass on the bias needs to be included in any precise analysis.

\begin{figure*}[htb]
\centering
\includegraphics[width=0.68\columnwidth]{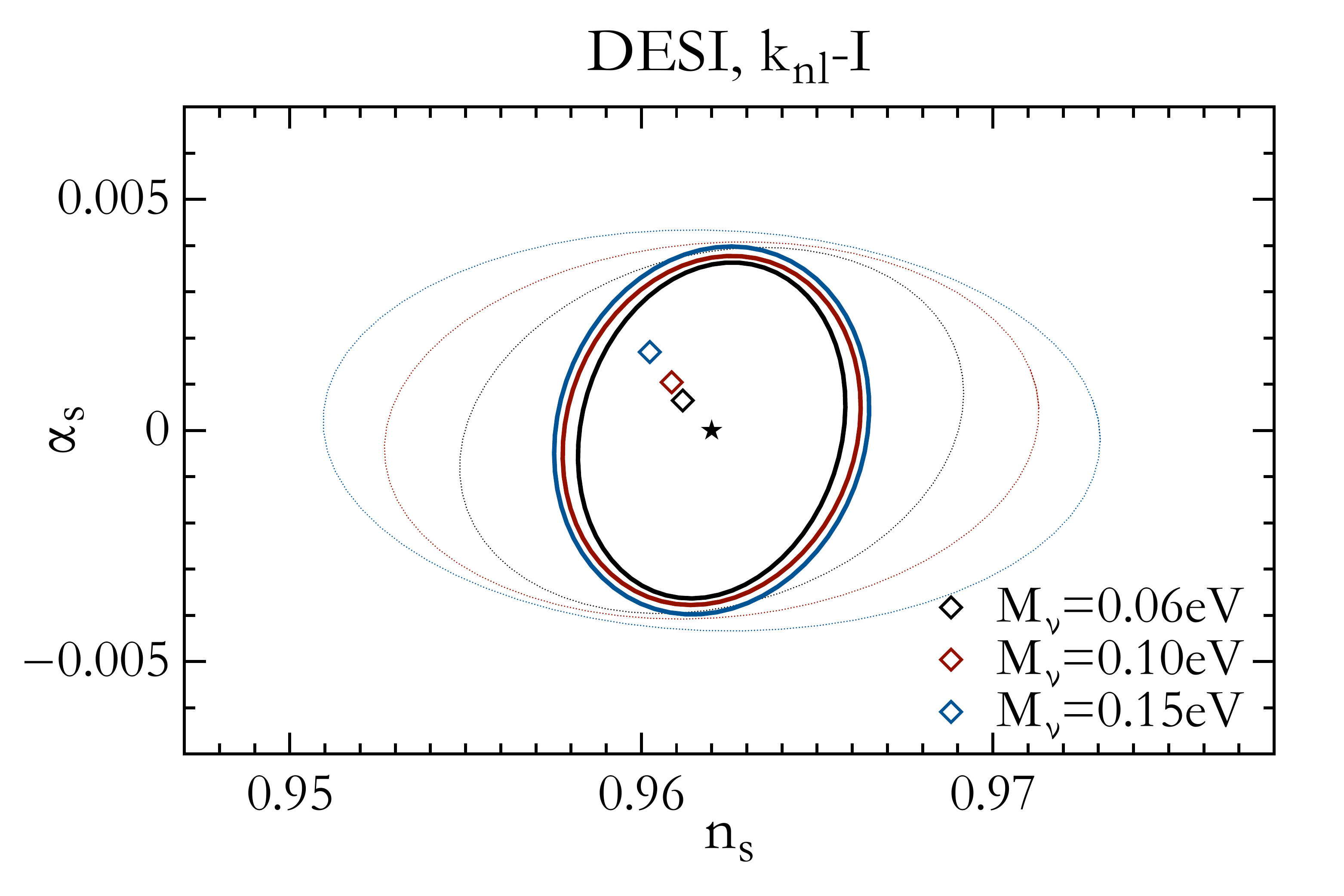}
\includegraphics[width=0.68\columnwidth]{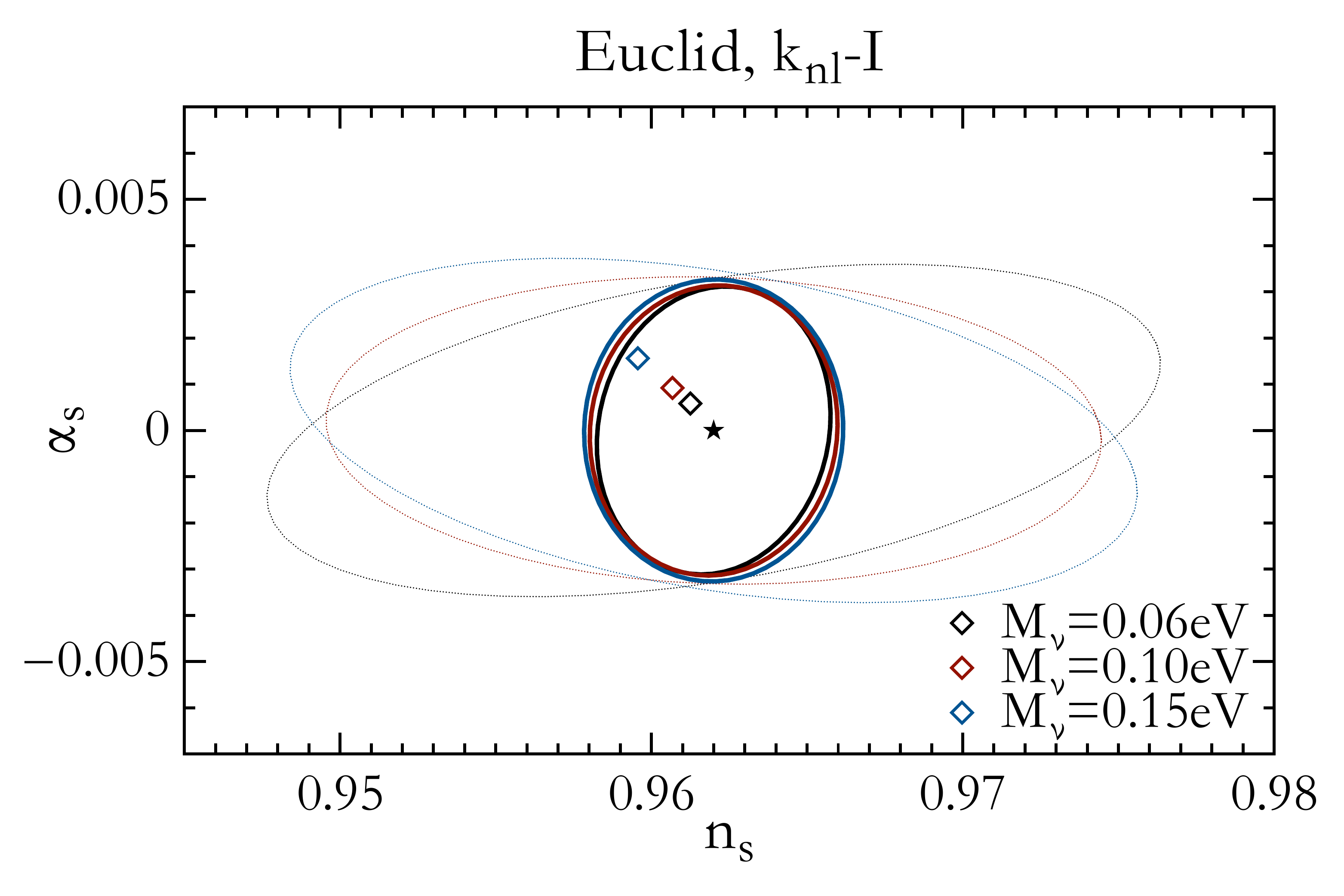}
\includegraphics[width=0.68\columnwidth]{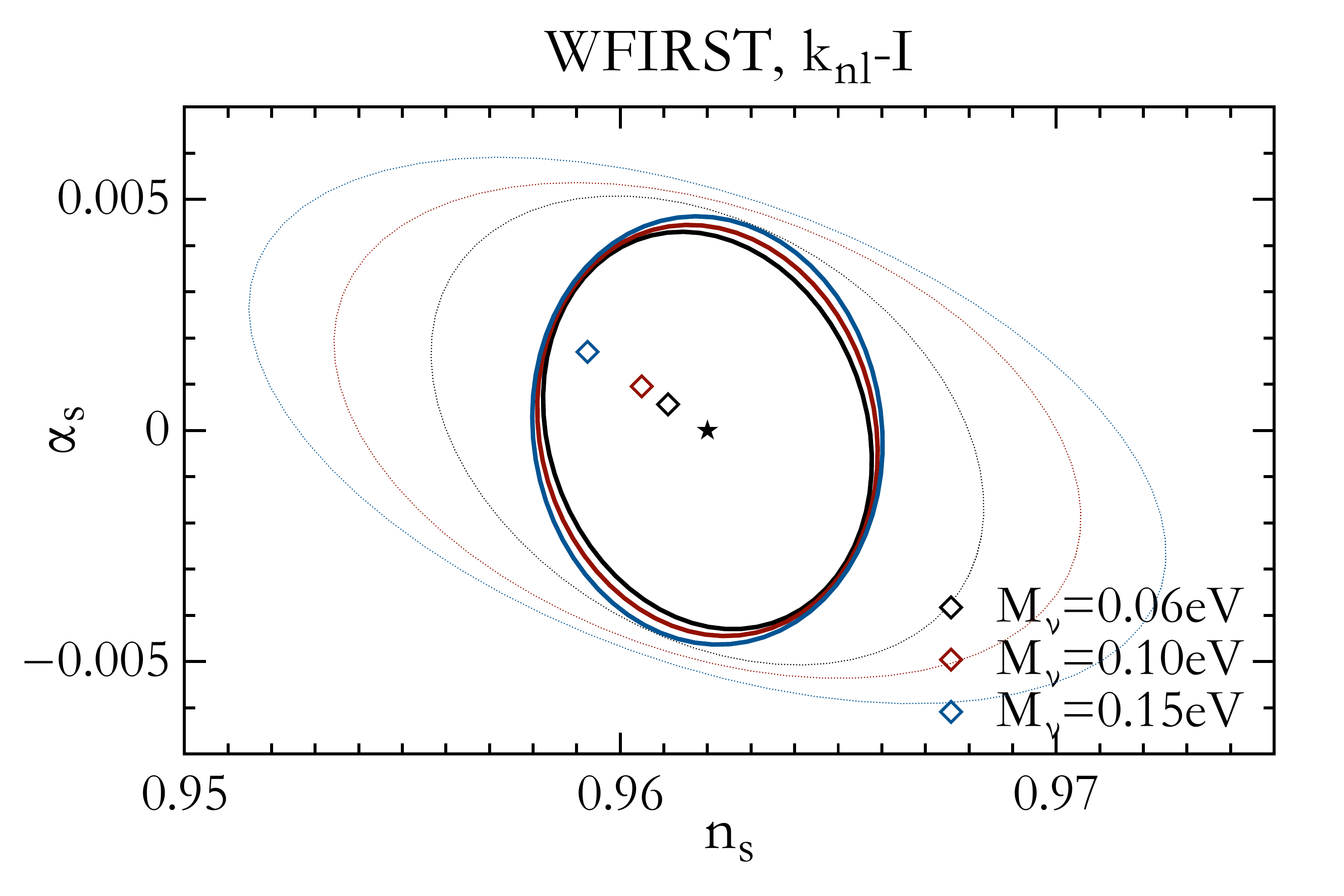}
\includegraphics[width=0.68\columnwidth]{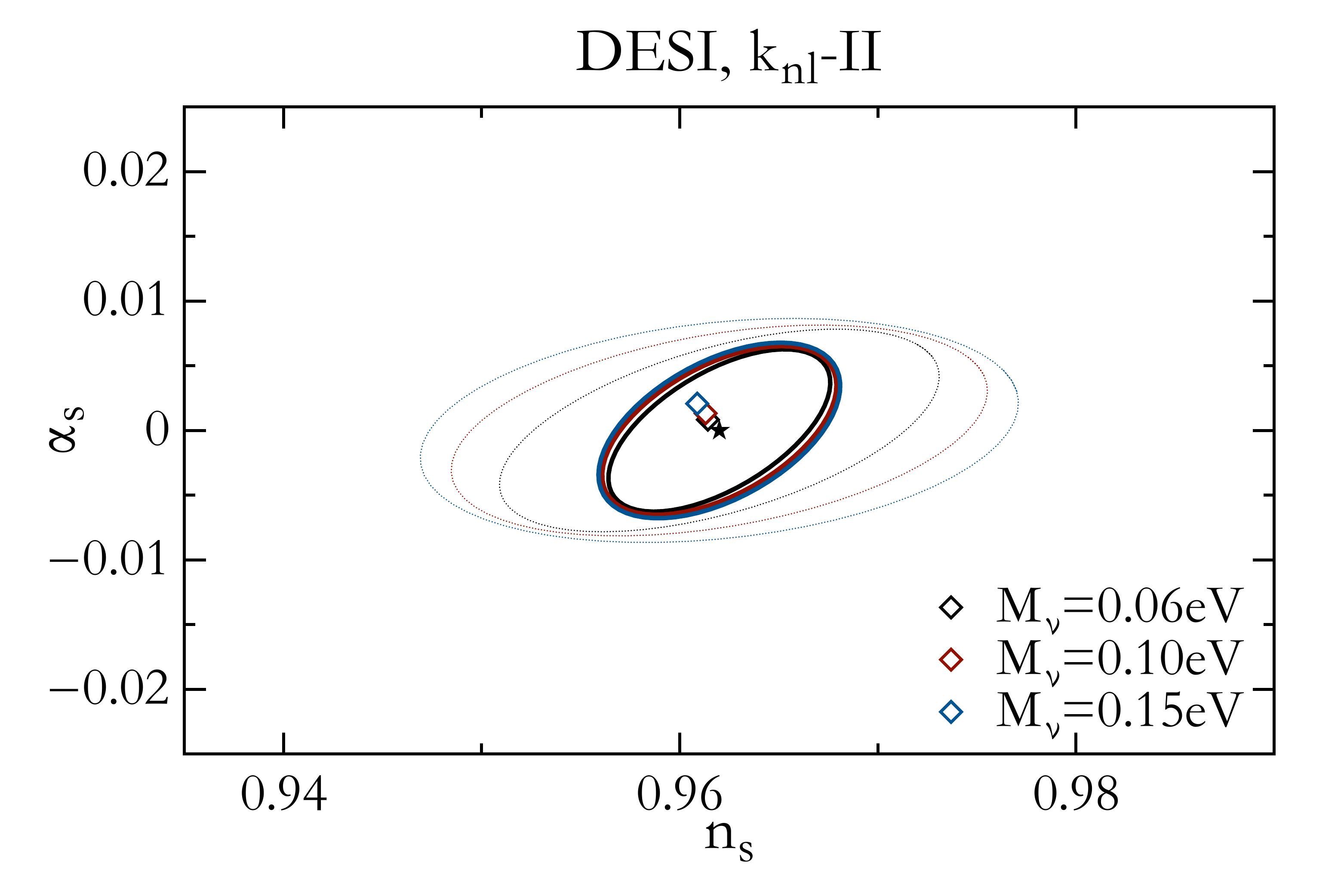}
\includegraphics[width=0.68\columnwidth]{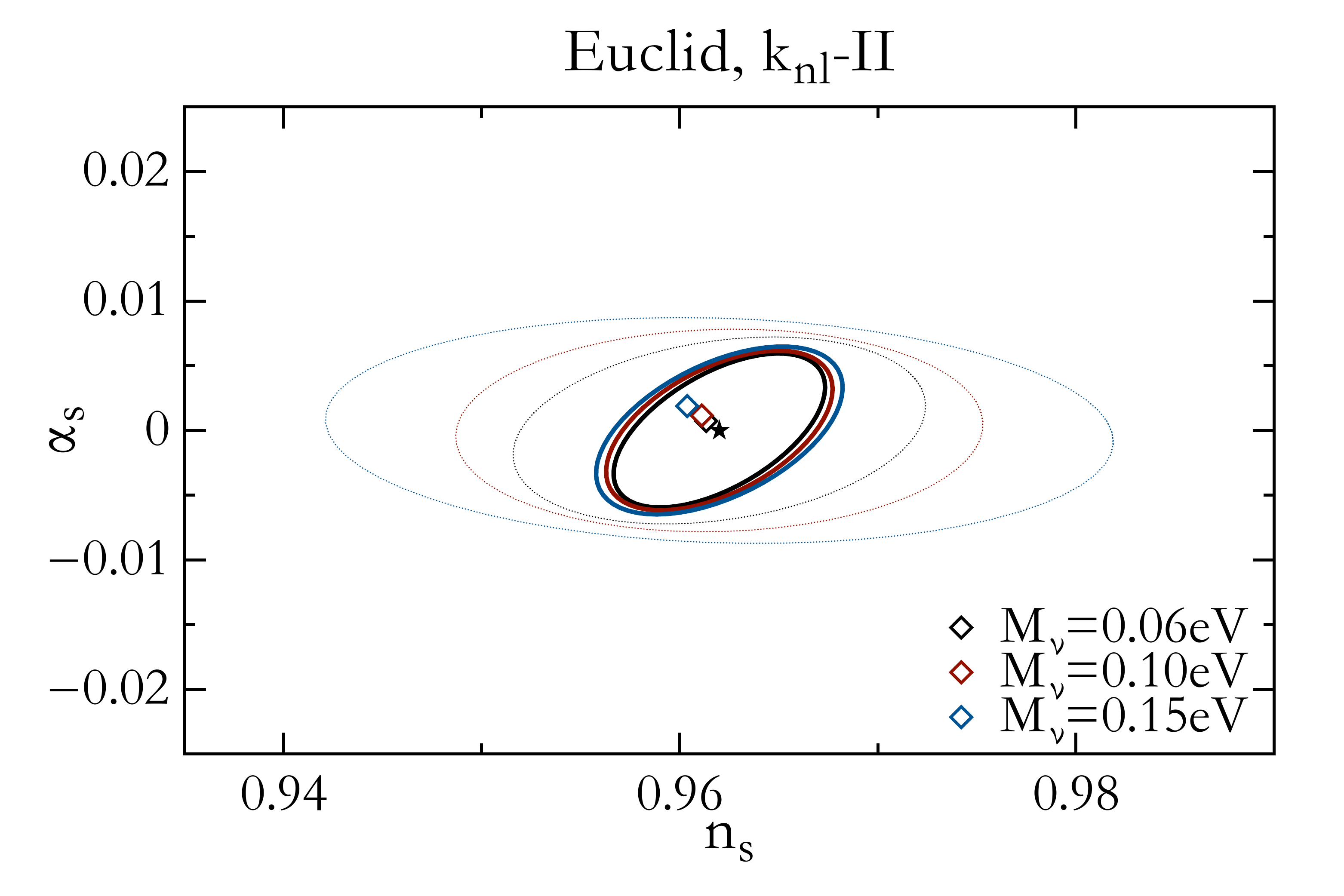}
\includegraphics[width=0.68\columnwidth]{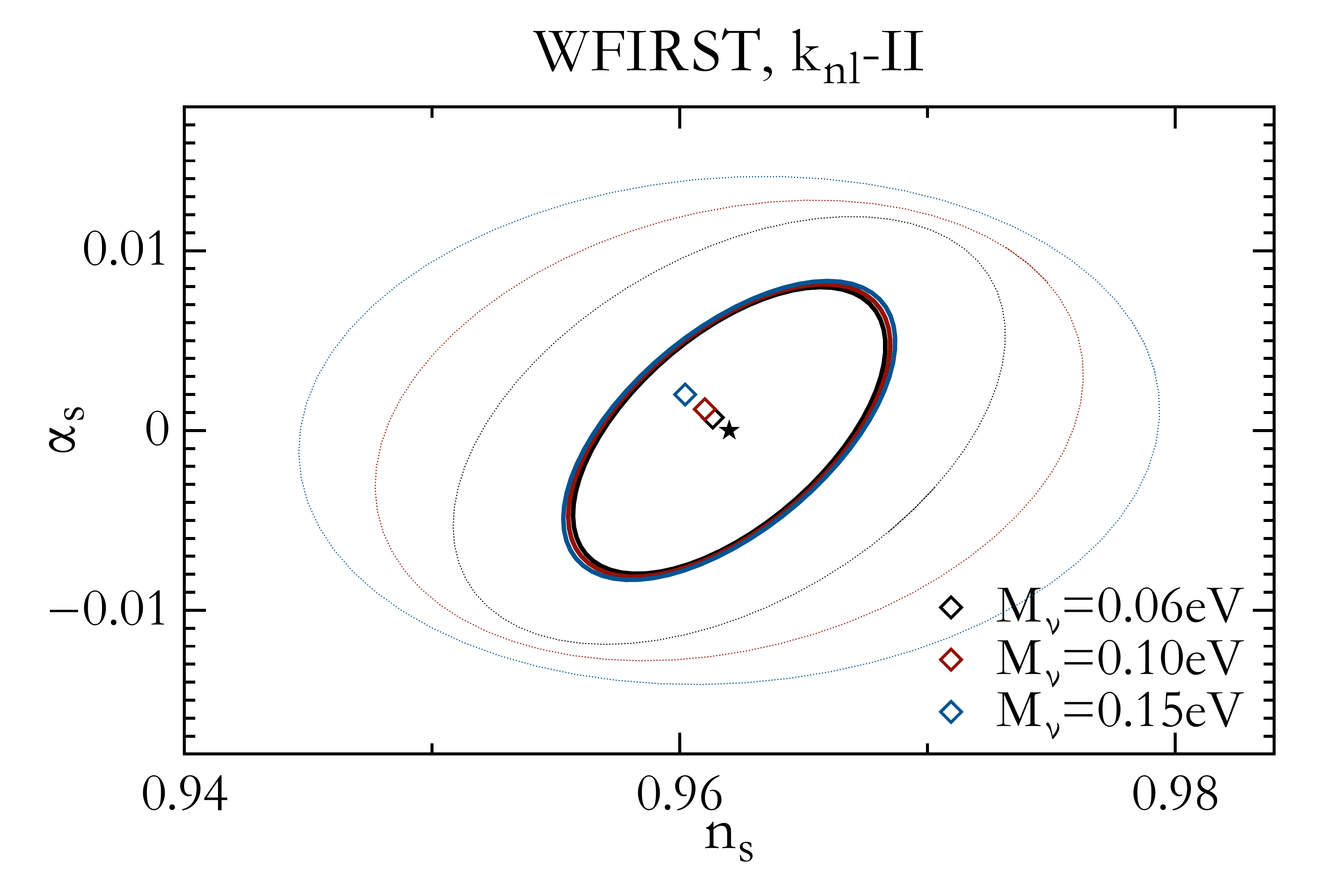}
\includegraphics[width=0.68\columnwidth]{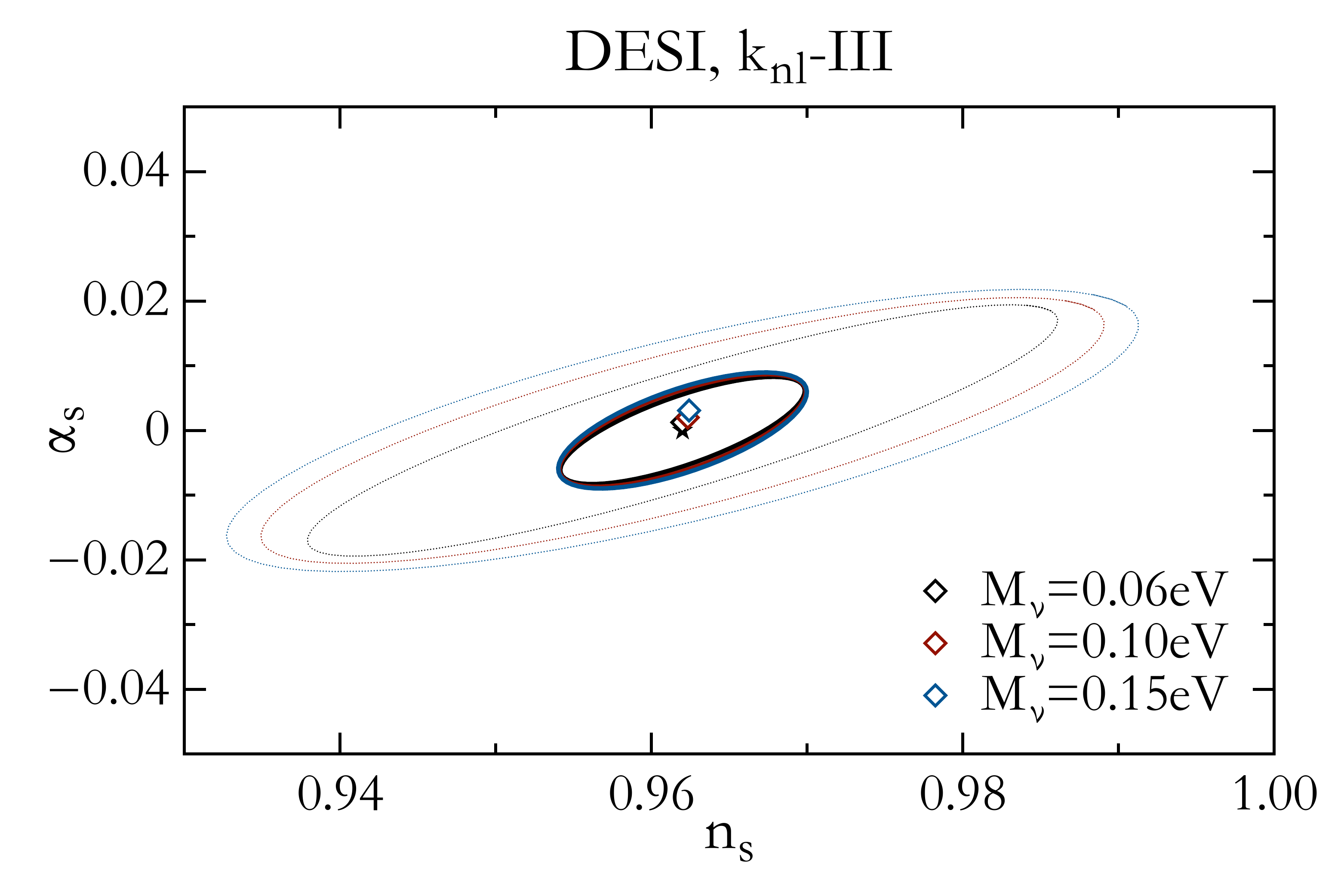}
\includegraphics[width=0.68\columnwidth]{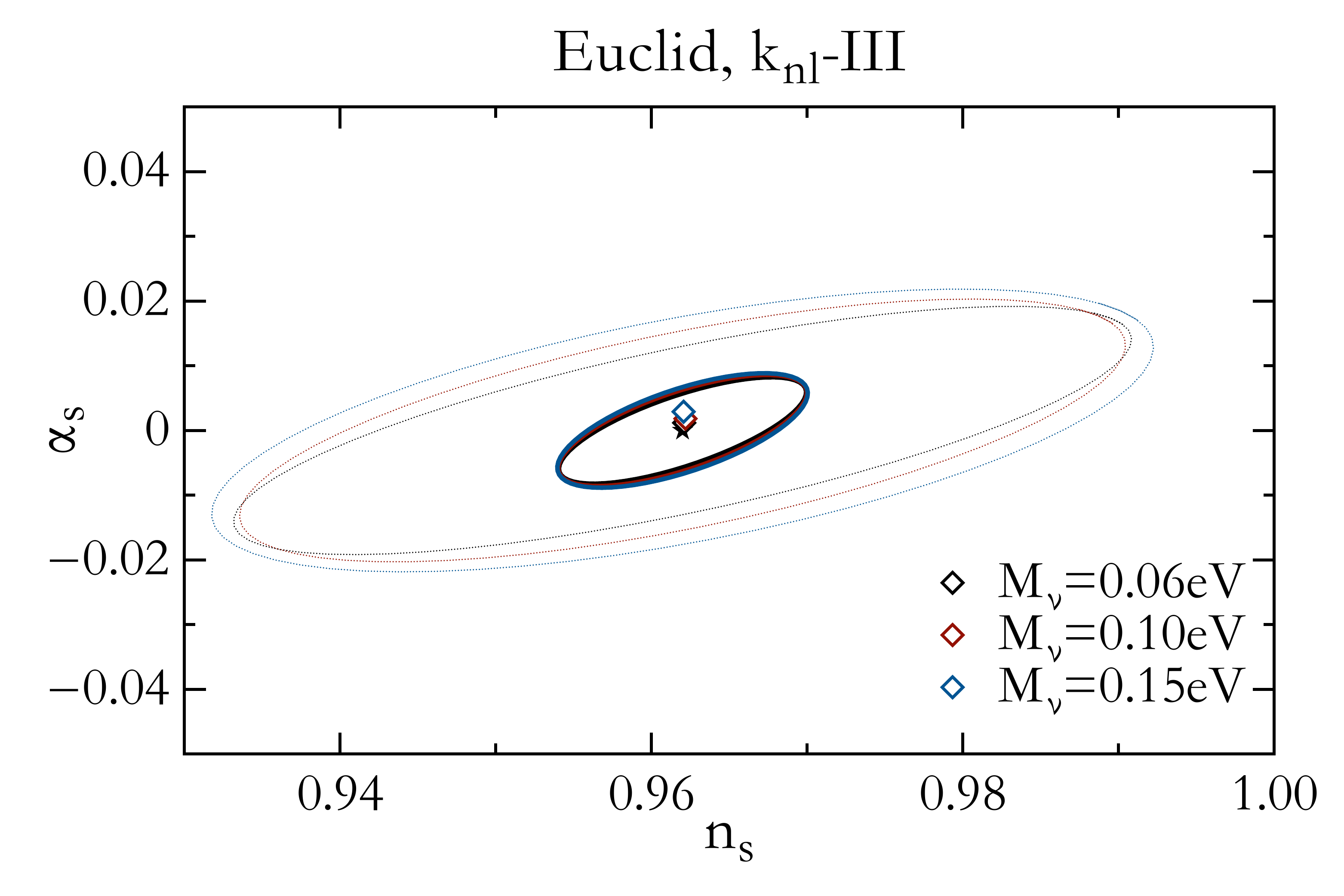}
\includegraphics[width=0.68\columnwidth]{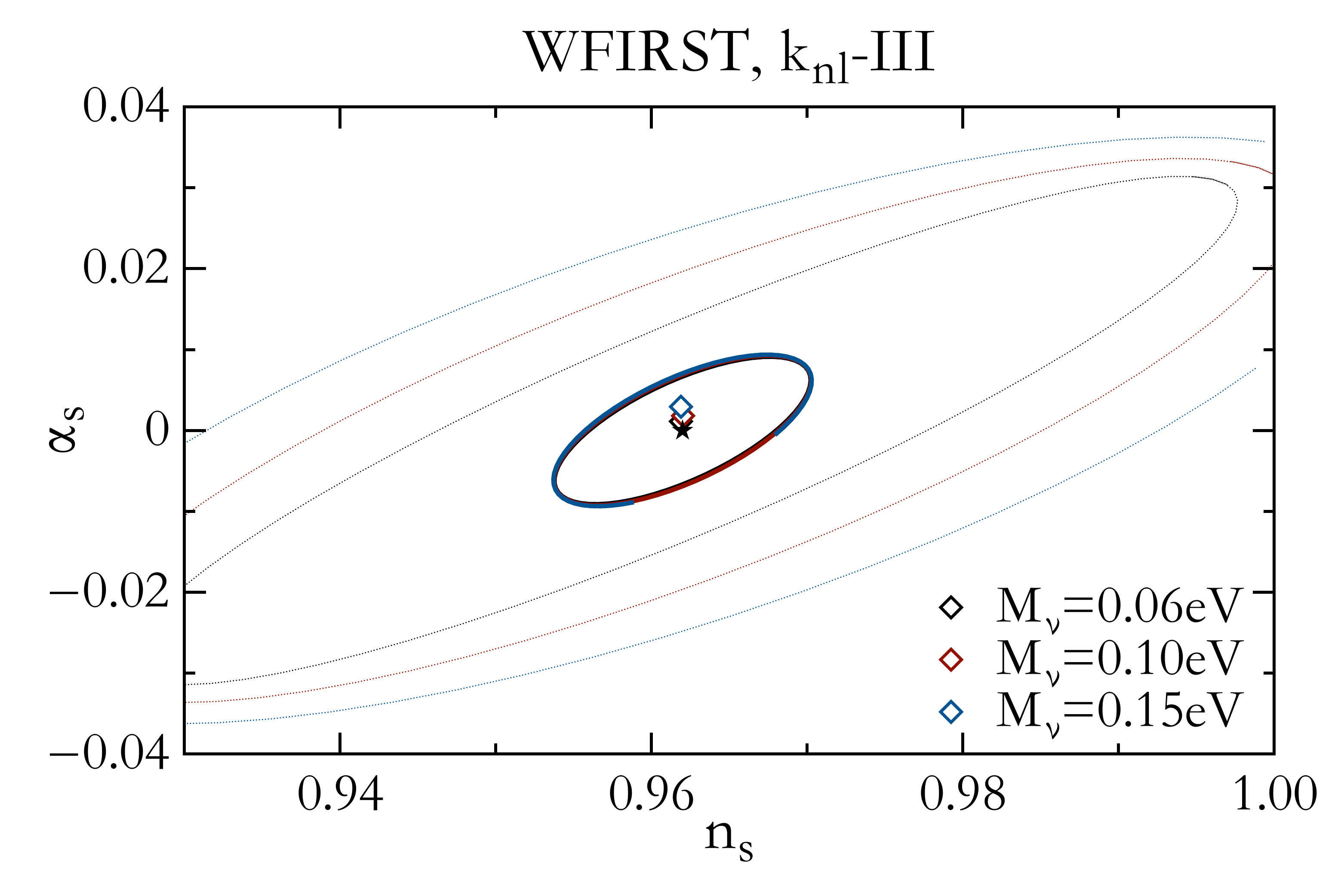}
\caption{As in Figure~\ref{fig:shift_nsas}, but when marginalizing over the  bias amplitude of three redshift bins.}
\label{fig:shift_nsas_bbb}
\end{figure*}

\begin{figure*}[htb]
\centering
\includegraphics[width=0.68\columnwidth]{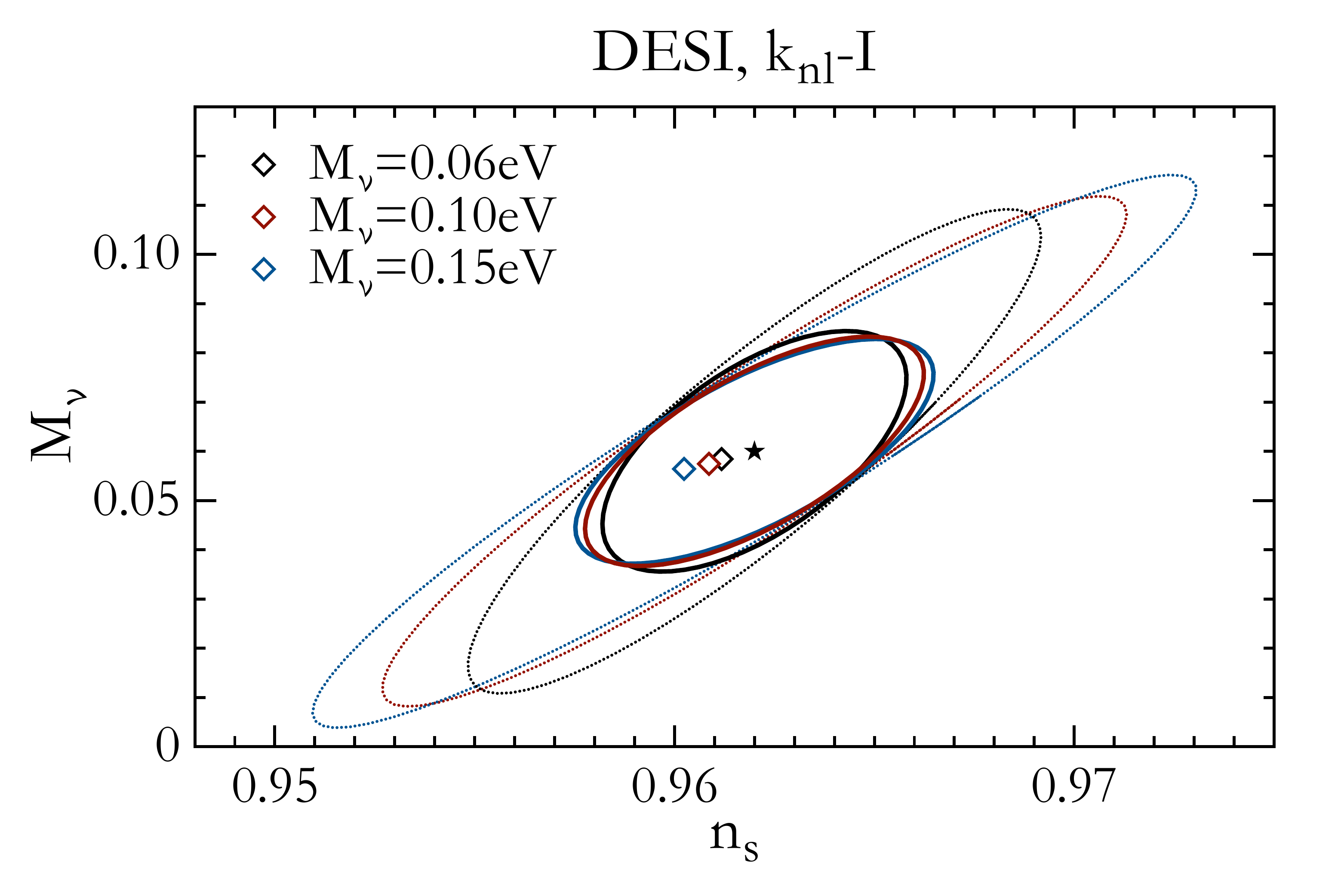}
\includegraphics[width=0.68\columnwidth]{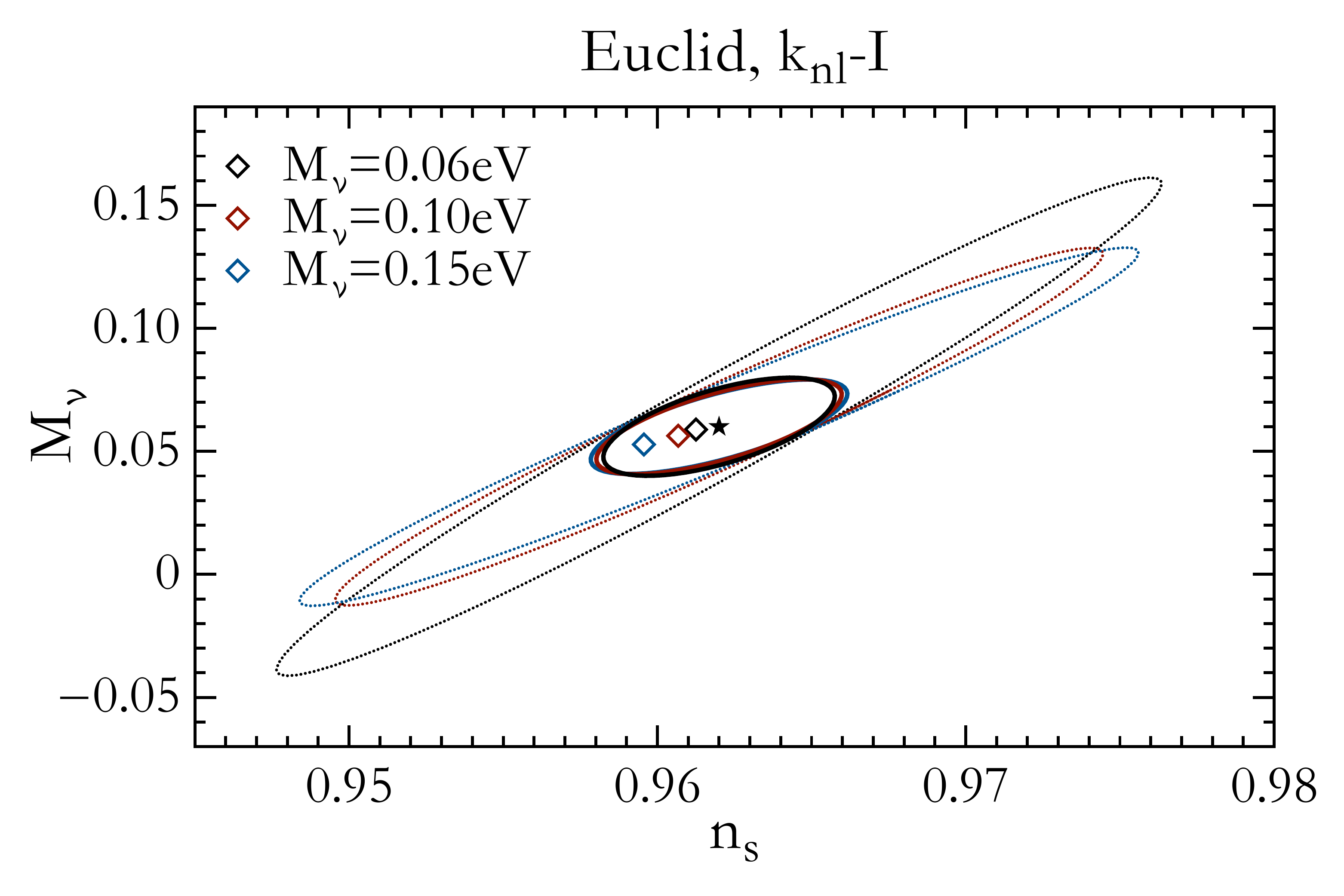}
\includegraphics[width=0.68\columnwidth]{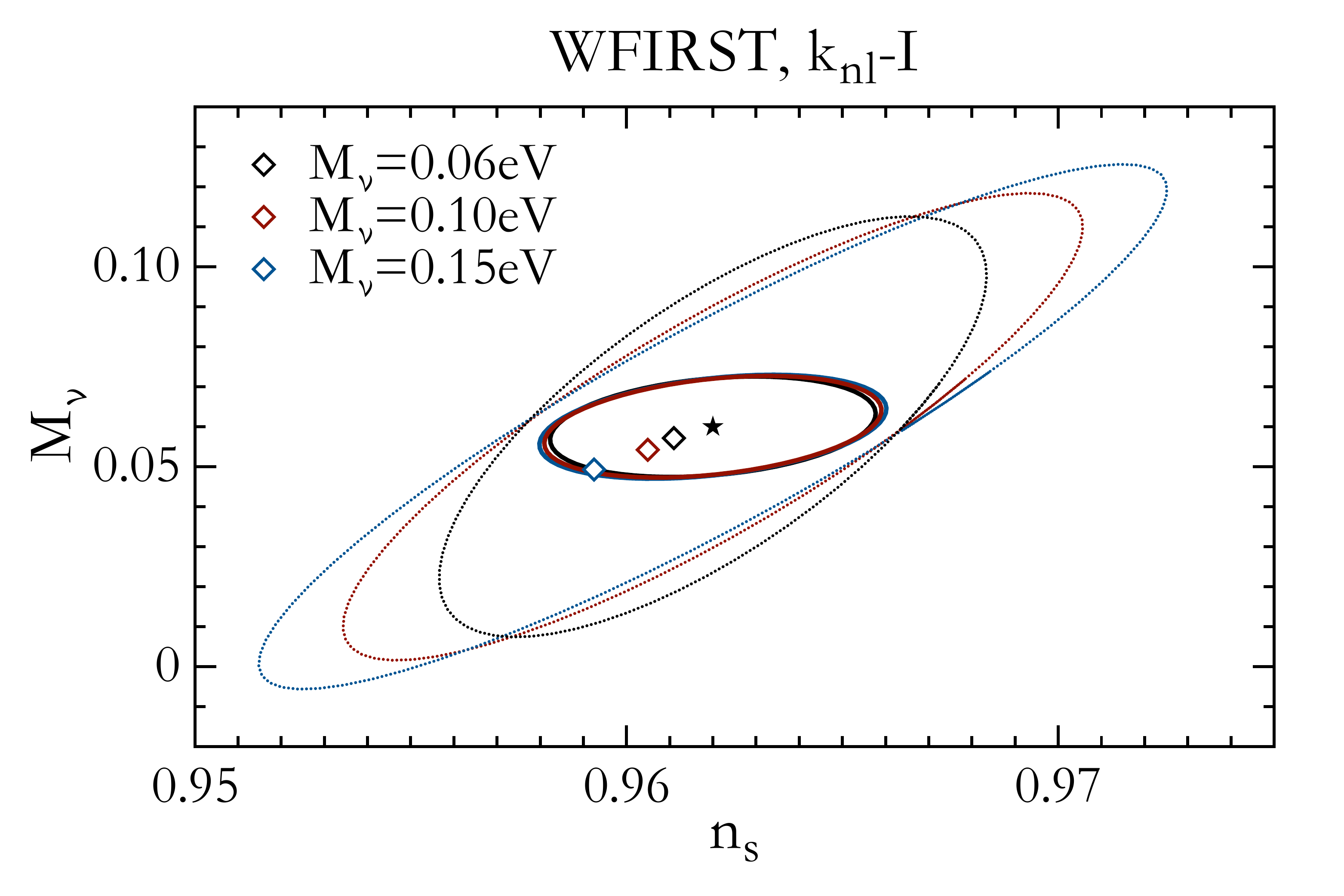}
\includegraphics[width=0.68\columnwidth]{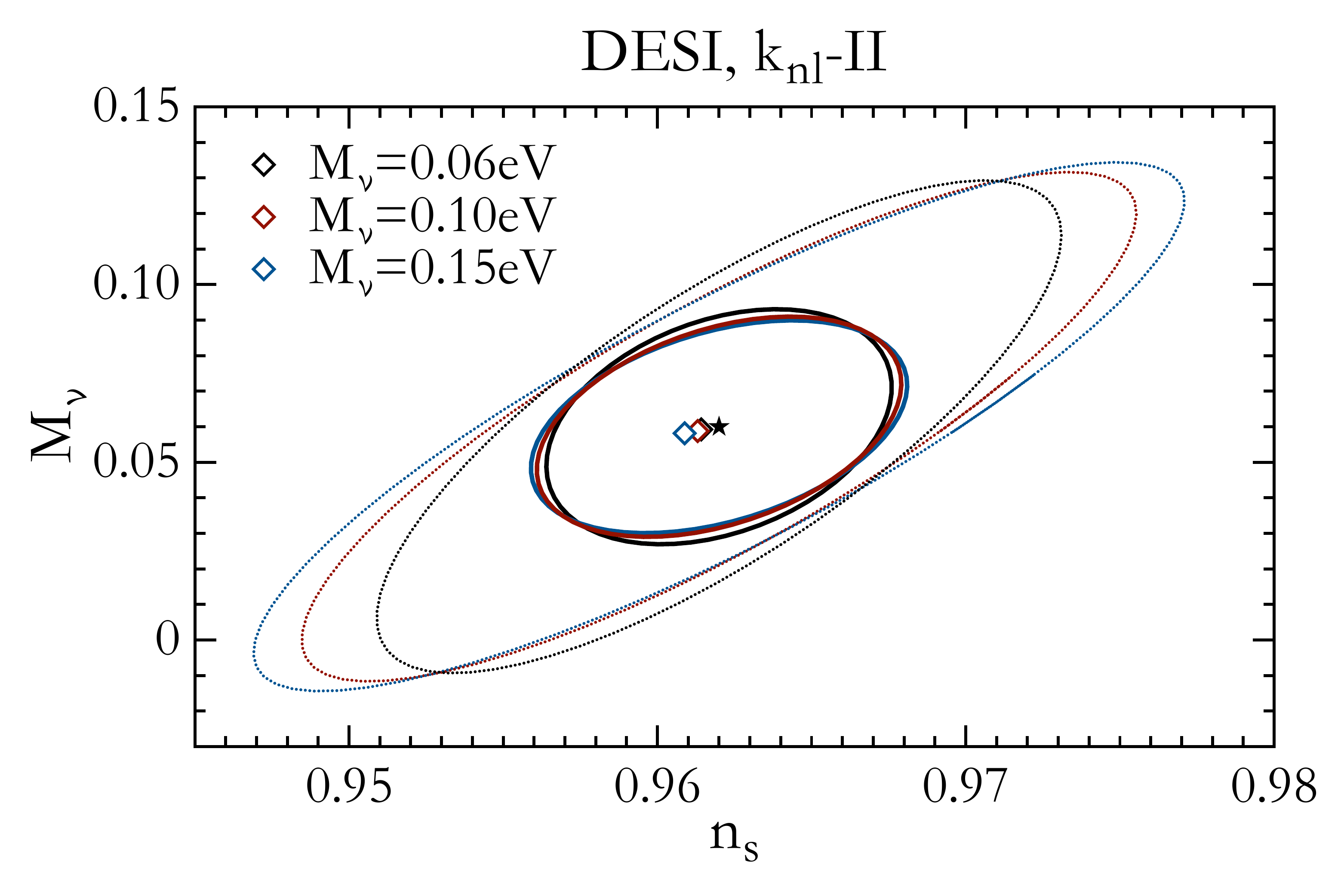}
\includegraphics[width=0.68\columnwidth]{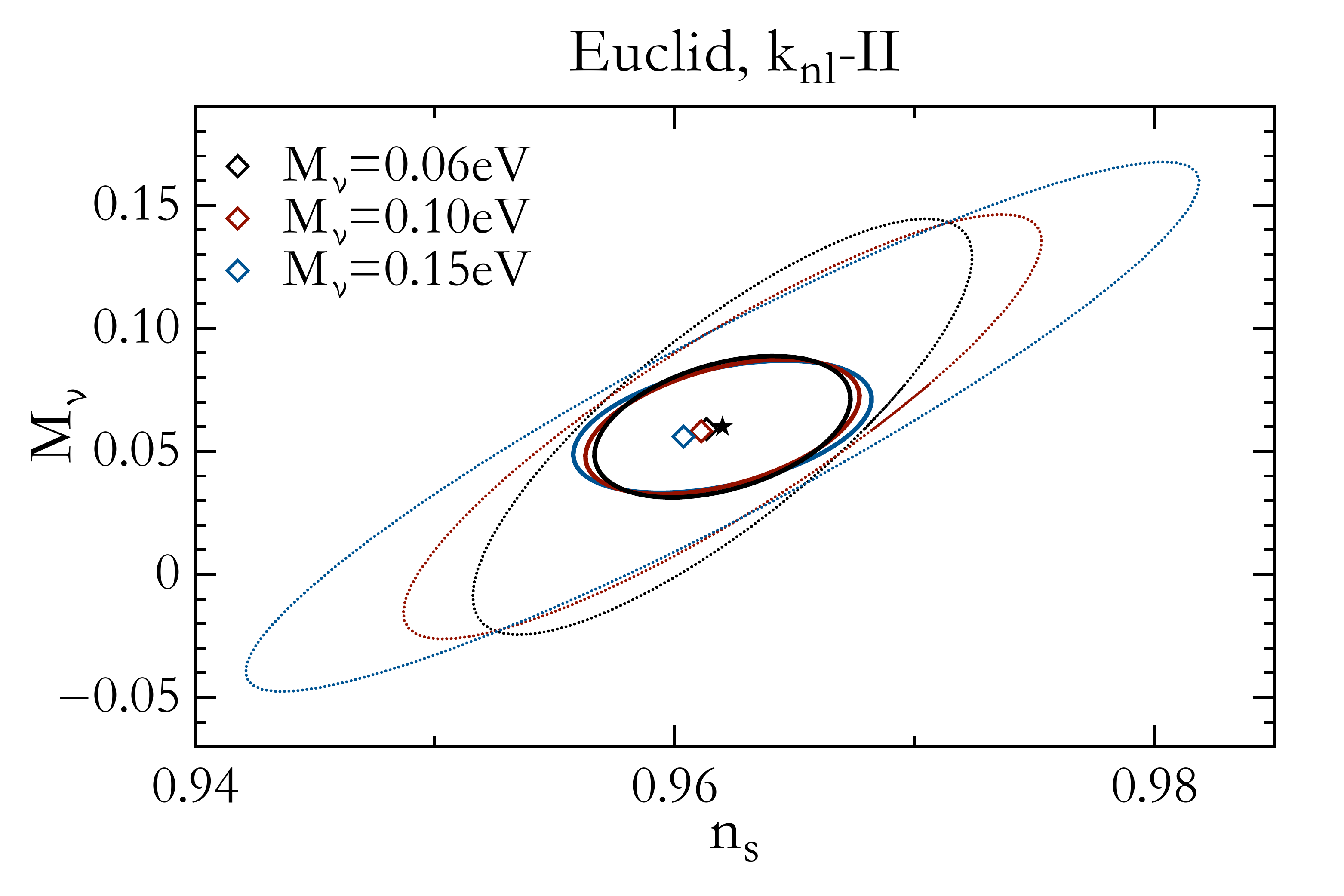}
\includegraphics[width=0.68\columnwidth]{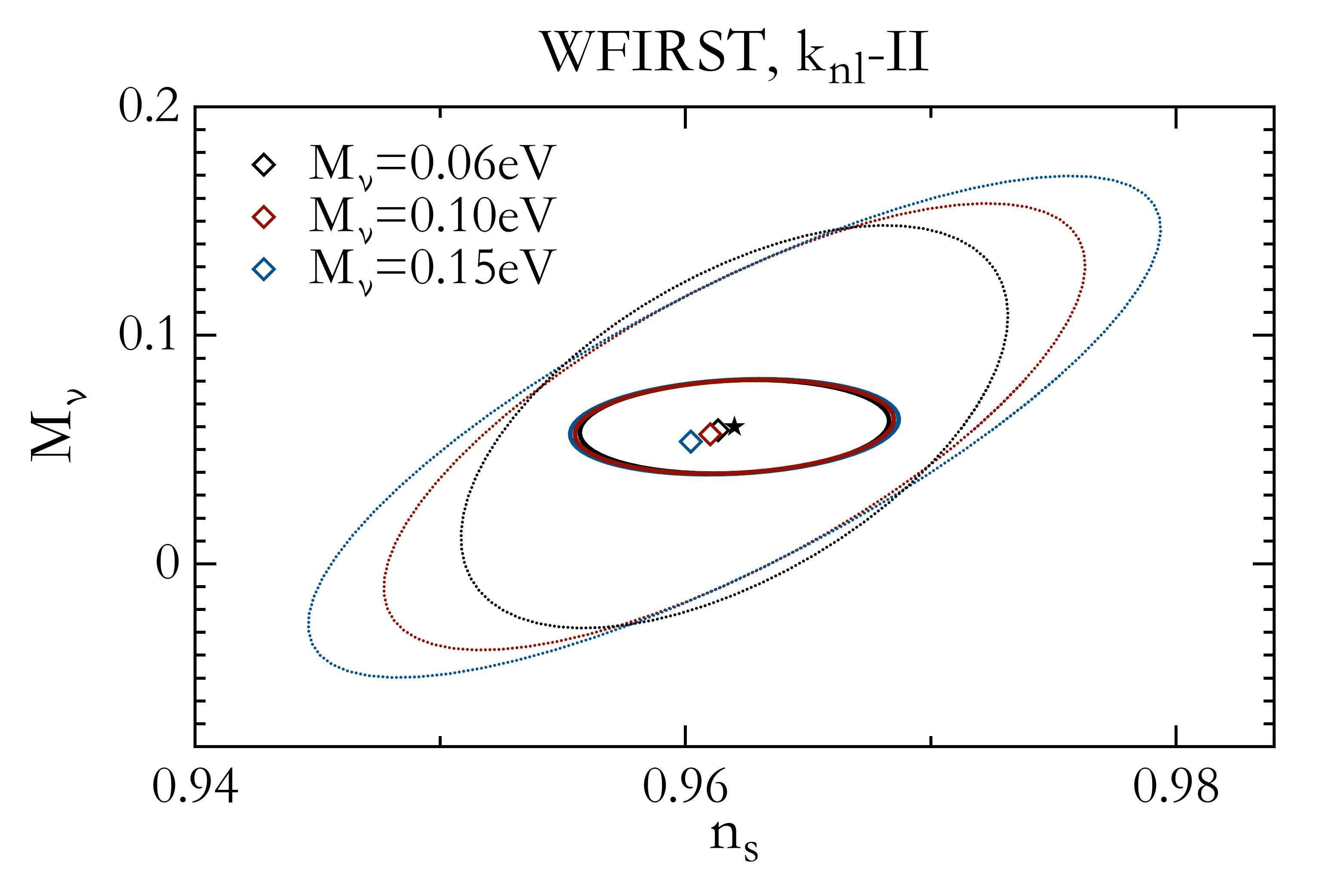}
\includegraphics[width=0.68\columnwidth]{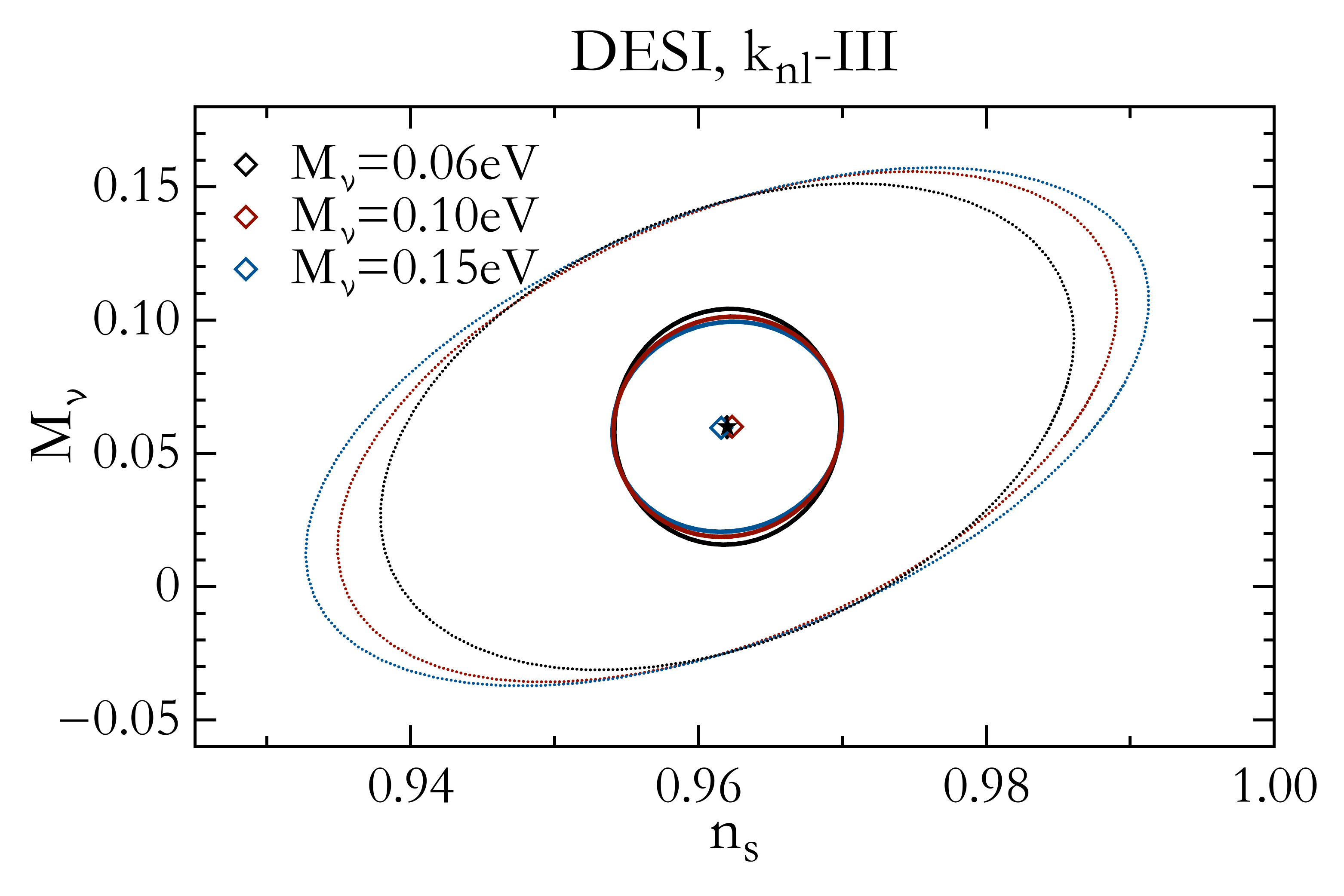}
\includegraphics[width=0.68\columnwidth]{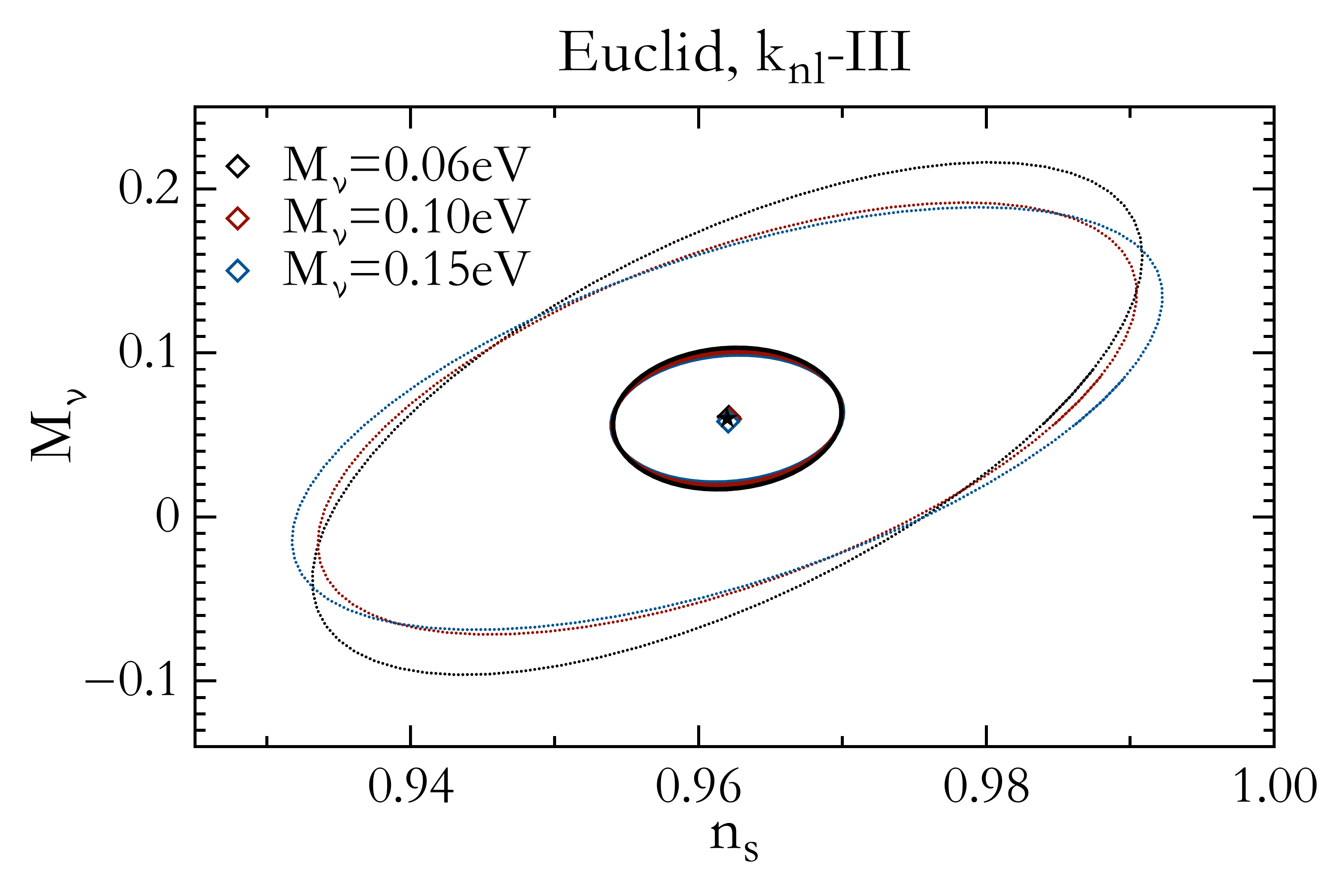}
\includegraphics[width=0.68\columnwidth]{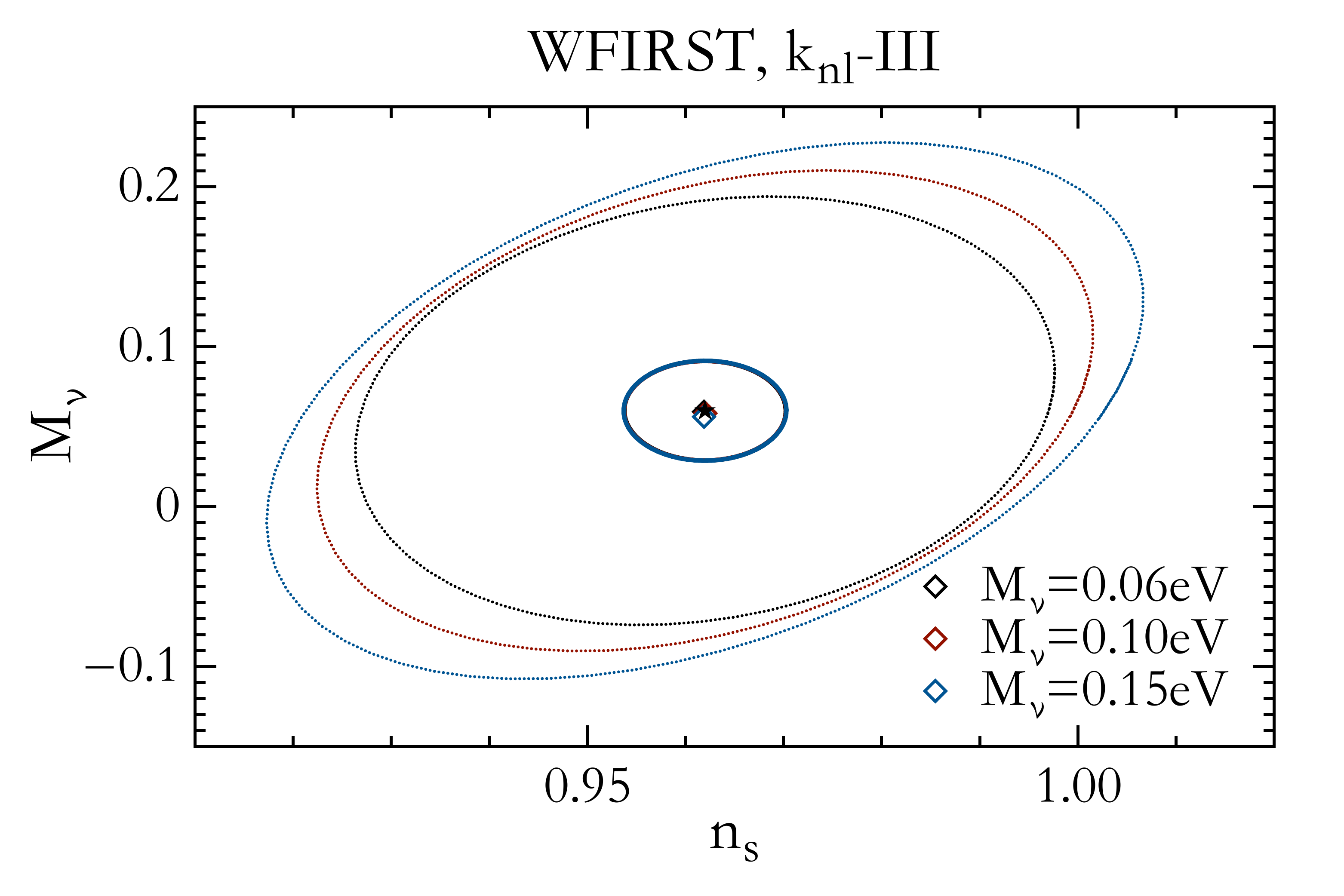}
\caption{As in Figure~\ref{fig:shift_nsMnu}, but when marginalizing over the  bias amplitude of three redshift bins.}
\label{fig:shift_nsMnu_bbb}
\end{figure*}

In Figures~\ref{fig:shift_nsas_bbb}-\ref{fig:shift_nsMnu_bbb} we present results for the case in which  marginalize over the bias  amplitude in three redshifts intervals independently, as explained above. In this case, as expected, the shifts are  significantly reduced compared to the predicted precision in the measurements. 

In fact  for the most conservative case, the $k_{nl}-III$ case,  shifts are totally negligible. But in the case of more aggressive approaches such as our $k_{nl}-I$  when including Planck priors, shifts will be a considerable fraction of the predicted sigma for $n_s$ and $M_\nu$. In a  realistic analysis, the accuracy (i.e.,  the systematic error budget) for each source of  systematic  error must obviously be well below the $1-\sigma$ statistical error, implying that this effect if possible should be corrected for in future high-precision cosmological analyses, regardless of  wether a more conservative of aggressive approach is used.

\section{Conclusions}
\label{sec:conclusions}
We have investigated  the magnitude of the systematic shift on interesting cosmological parameters induced by neglecting the effect of massive neutrinos on halo bias. 
Understanding and modelling a survey's tracers bias is one of the main difficulties to be faced to measure the signature of neutrino masses from future redshift surveys. As tracers are hosted in dark matter halos, one of the crucial steps  is to model correctly the bias of the halo field or halo bias.

Since massive neutrinos  induce a scale-dependent growth of perturbations, they induce a scale-dependent modification to the halo bias compared to a case where neutrinos are massless. It is well known that the constant, linear bias model is 
incorrect and cannot be  employed lightly in the analysis of present and future surveys. However the additional effect  introduced by neutrino masses is usually neglected; here we have quantified the effect of this approximation. 

We confirm that for current survey this is not an issue.
We computed results for a variety of planned forthcoming galaxy surveys such as DESI and the space missions Euclid and WFIRST, as representative of wide and deep surveys, but the gist of our results will be valid for most future surveys.
We present forecasts for different cases of neutrino masses and power spectrum modelings, in particular different values of the minimum scale used for the analyses.

Our results show that the shifts in units of predicted measurement precisions are very dependent on the case considered, the marginalization over one or many bias parameters, and the maximum wavenumber used. In general, however, apart from the most conservative  cases, the shifts will  range between tens of percent to slightly more than a $1-\sigma$ on quantities like the  power spectrum spectral index and the neutrino mass itself.
In a  realistic analysis, the accuracy (i.e.,  the systematic error budget) for each source of  systematic  error must obviously be well below the $1-\sigma$ statistical error, implying that this effect should be modelled, in forthcoming, high-precision redshift surveys. This is especially true if one wants to include mildly non-linear scales to extract the maximum amount of information possible.

We argue that  this can be  done easily  and at no additional computational cost  by using the bias computed with respect of the CDM ($b_{\rm cc}$) only instead of the total matter ($b_{\rm mm}$), as shown in~\cite{paco14,Castorina:2013wga, loverde}. This  quantity can be studied and modelled with the help of N-body simulations (as we have done here, also providing accurate  fitting functions); it has the advantage that it is universal and its scale dependence does not depend on the value of the neutrino mass, thus reducing the computational costs. 

Thus  we envision the recipe to account for the neutrino mass effect on halo bias to be as follows. The $b_{\rm cc}$ quantity is obtained from a suite of N-body simulations and bias values at finer sampling than offered by the simulations can be obtained by interpolation or developing an emulator.
Since  any realistic analysis of data require the bias with respect to the total matter, $b_{mm}$, the ratio  $b_{\rm cc}/b_{\rm mm}$ can be computed analytically as we have discussed in sec.~\ref{sec:bias}. After the initial investment in modelling $b_{\rm cc}$ this approach does not increase significantly the  computational cost of the analysis, but remove efficiently  the systematic shift on the recovered cosmological parameters.
We hope that this proposed approach will be useful to improve the robustness of cosmological results from forthcoming surveys.

\vspace{0.3cm}

{\bf Acknowledgments} \\
We would like to thank Emanuele Castorina, Donghui Jeong and Anthony Pullen for useful discussions.
AR has received funding from the People Programme (Marie Curie Actions) of the European Union H2020 Programme under REA grant agreement number 706896 (COSMOFLAGS). Funding for this work was partially provided by the Spanish MINECO under MDM-2014-0369 of ICCUB (Unidad de Excelencia ``Maria de Maeztu'') and by MINECO grant  AYA2014-58747-P AEI/FEDER, UE.
FVN is supported by the Simons Foundation.

\appendix
\section{Bias fits}
\label{sec:appendix}
We provide  smooth fitting functions to the scale dependent bias as a function of  scale (i.e., wavenumber $k$) and  redshift.
We begin presenting the fit to the bias with respect to cold dark matter $b_{\rm cc}$   (for the $M_{\nu}=0$ case) as its shape does not depend on neutrino masses as shown in~\cite{Castorina:2013wga}.
We use a fourth order polynomial function for each redshift:
\begin{equation}
\label{eq:bcc}
b_{cc}(k,z) = b_1(z) + b_2(z) k^2 + b_3(z) k^3 + b_4(z) k^4 \, .
\end{equation}
In Table~\ref{tab:fitbcc} we provide the values of the fit's coefficients for selected values of $z$ used in this work, and Figure~\ref{tab:fitbcc} shows the bias from the simulations and the fit adopted\footnote{While in principle for symmetry considerations the bias should only have  even powers of $k$, here we consider a purely phenomenological fit over a reduced range of scales and thus we use a generic polynomial.}. For a practical application we envision a finer redshift sampling being explored with simulations and  an interpolation/emulator procedure for redshifts  falling between the sampled values.

\begin{table}[htb]
	\begin{tabular}{|c|c|c|c|c|}
		\hline
		Redshift & $b_1$ & $b_2$ & $b_3$ & $b_4$  \\
		\hline
		0 & 1.14 & -1.23 & -2.19 & 4.77 \\
		\hline
		1 & 2.13 & 3.37 & -12.94 & 11.6 \\
		\hline 
		2 & 3.85 & 22.06 & -60.5 & 49.2 \\
		\hline
		3 & 6.46 & 94.02 & -262.05 & 227.27 \\
		\hline       
	\end{tabular}
	\caption{Coefficients of the $b_{cc}$ fit of Equation~\ref{eq:bcc}.}
	\label{tab:fitbcc}
\end{table}

How $b_{mm}$ is obtained from $b_{cc}$ is described in Equation~\eqref{eq:5} of Section~\ref{sec:bias}.

\begin{figure*}
\centering
\includegraphics[width=0.95\columnwidth]{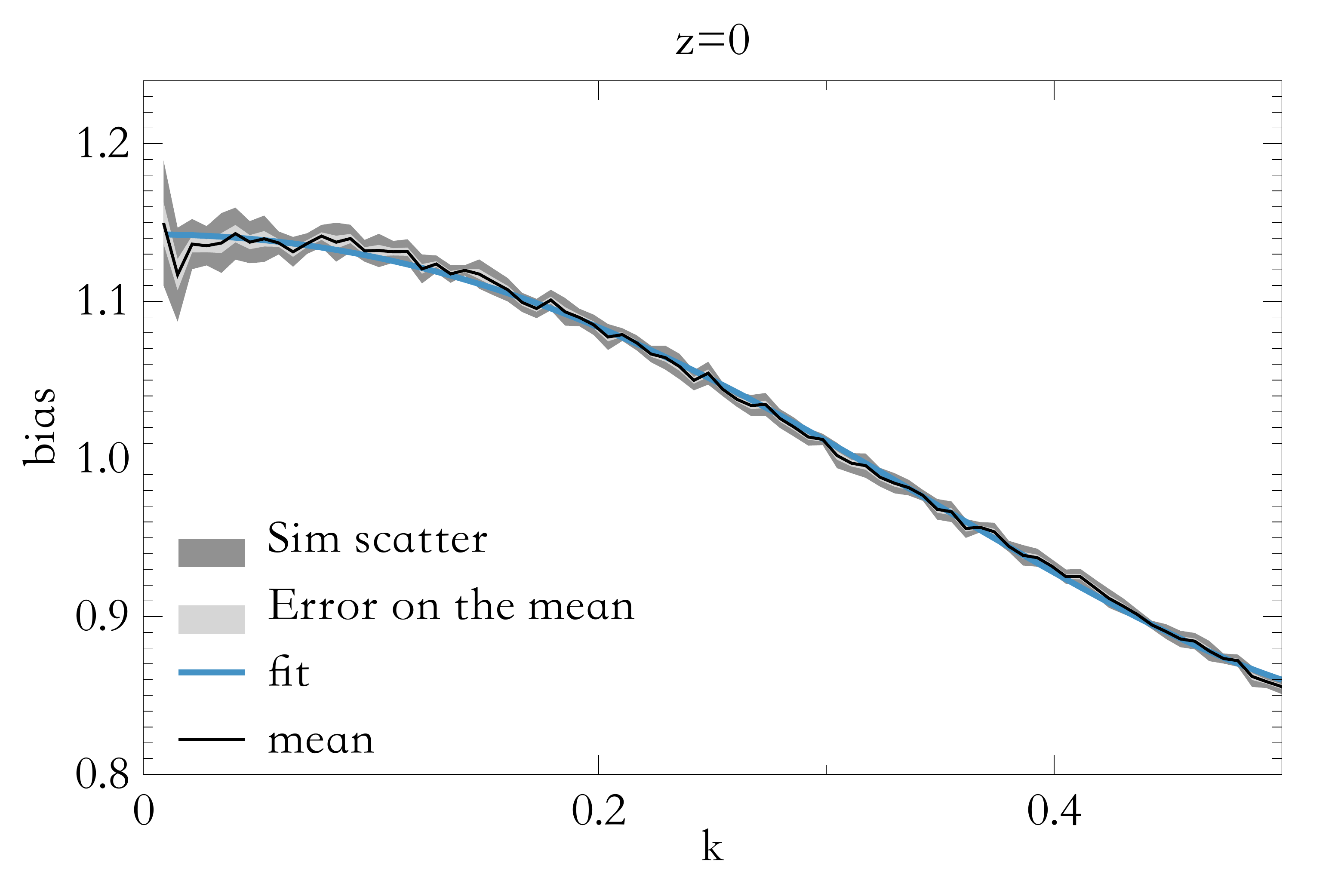}
\includegraphics[width=0.95\columnwidth]{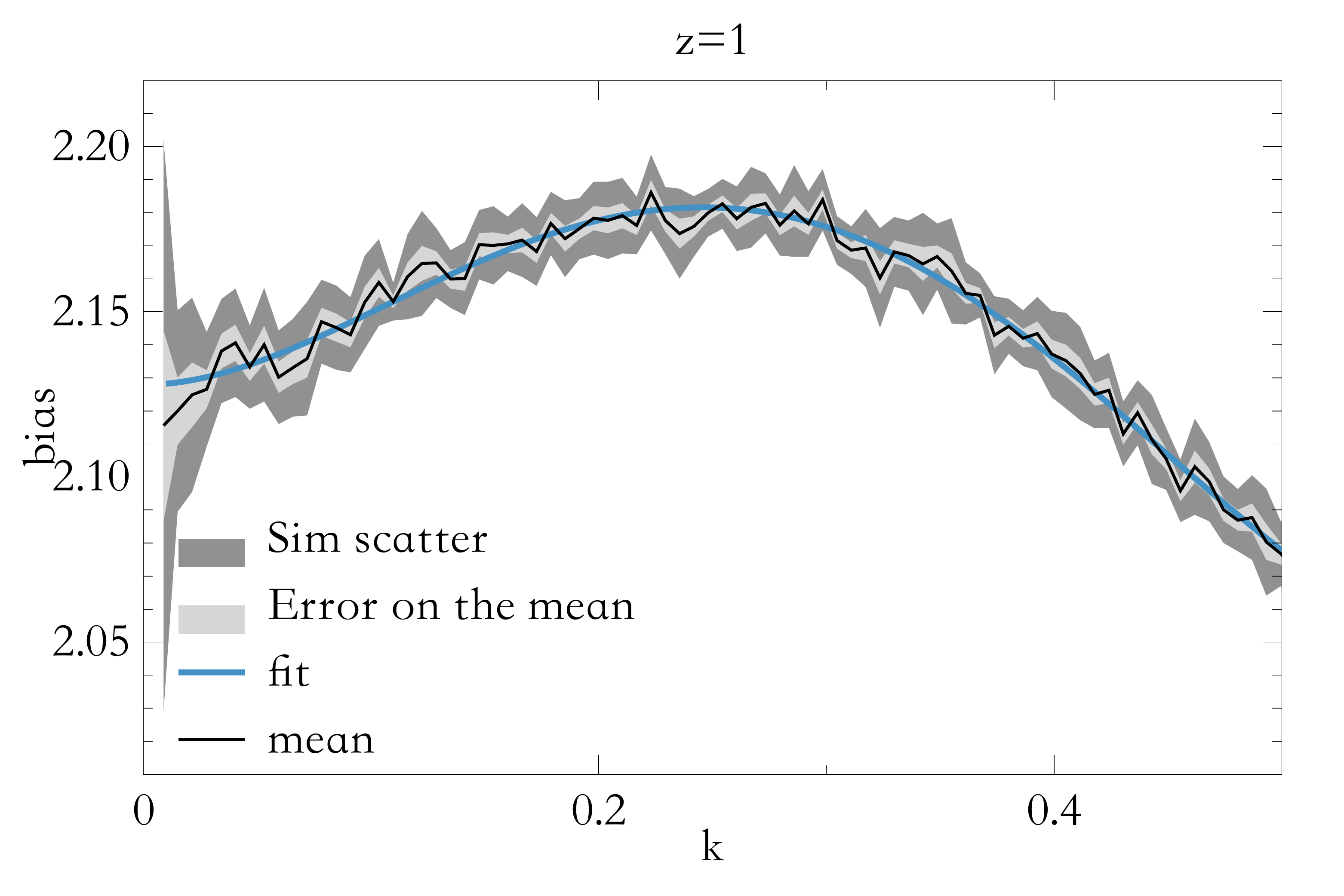}
\includegraphics[width=0.95\columnwidth]{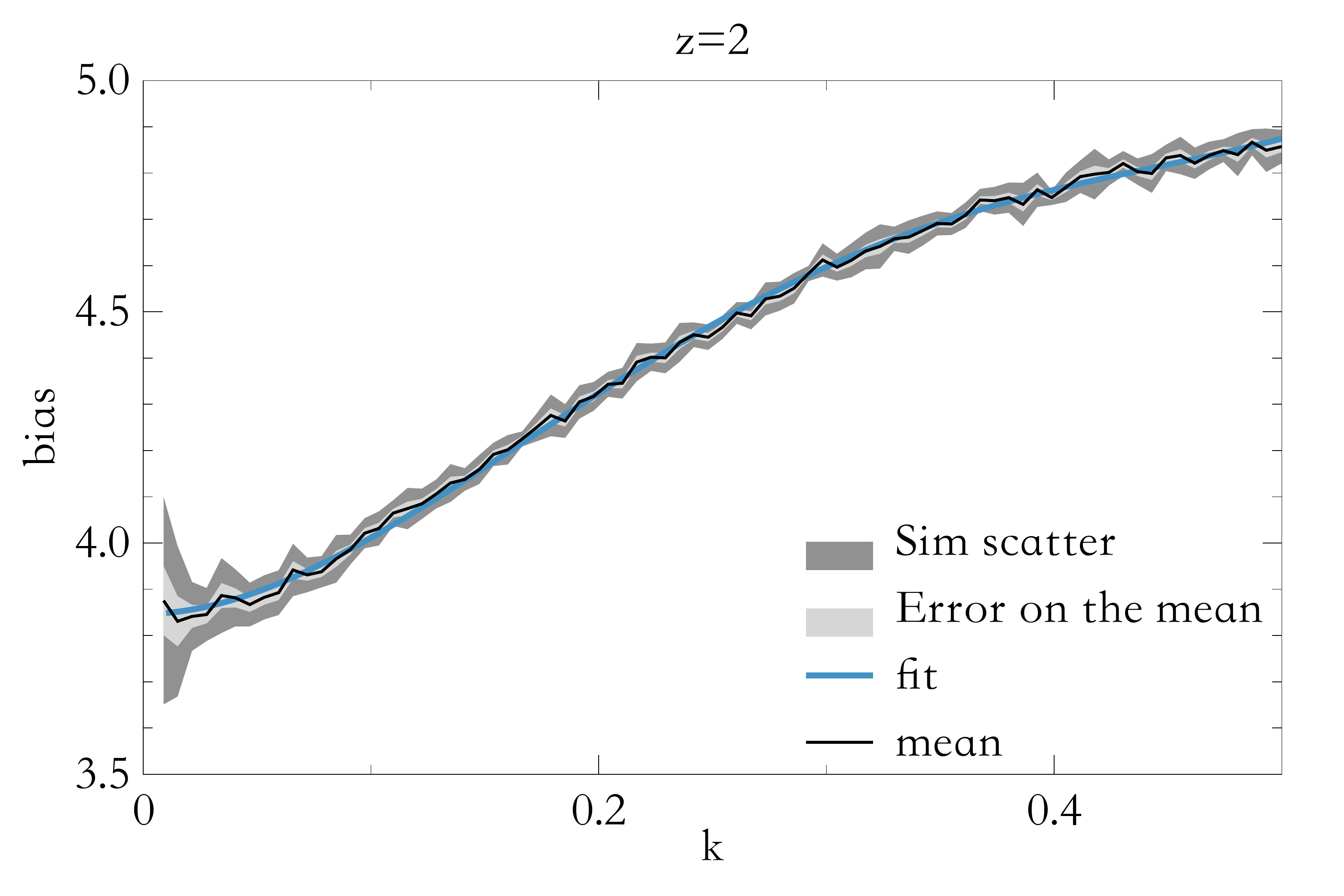}
\includegraphics[width=0.95\columnwidth]{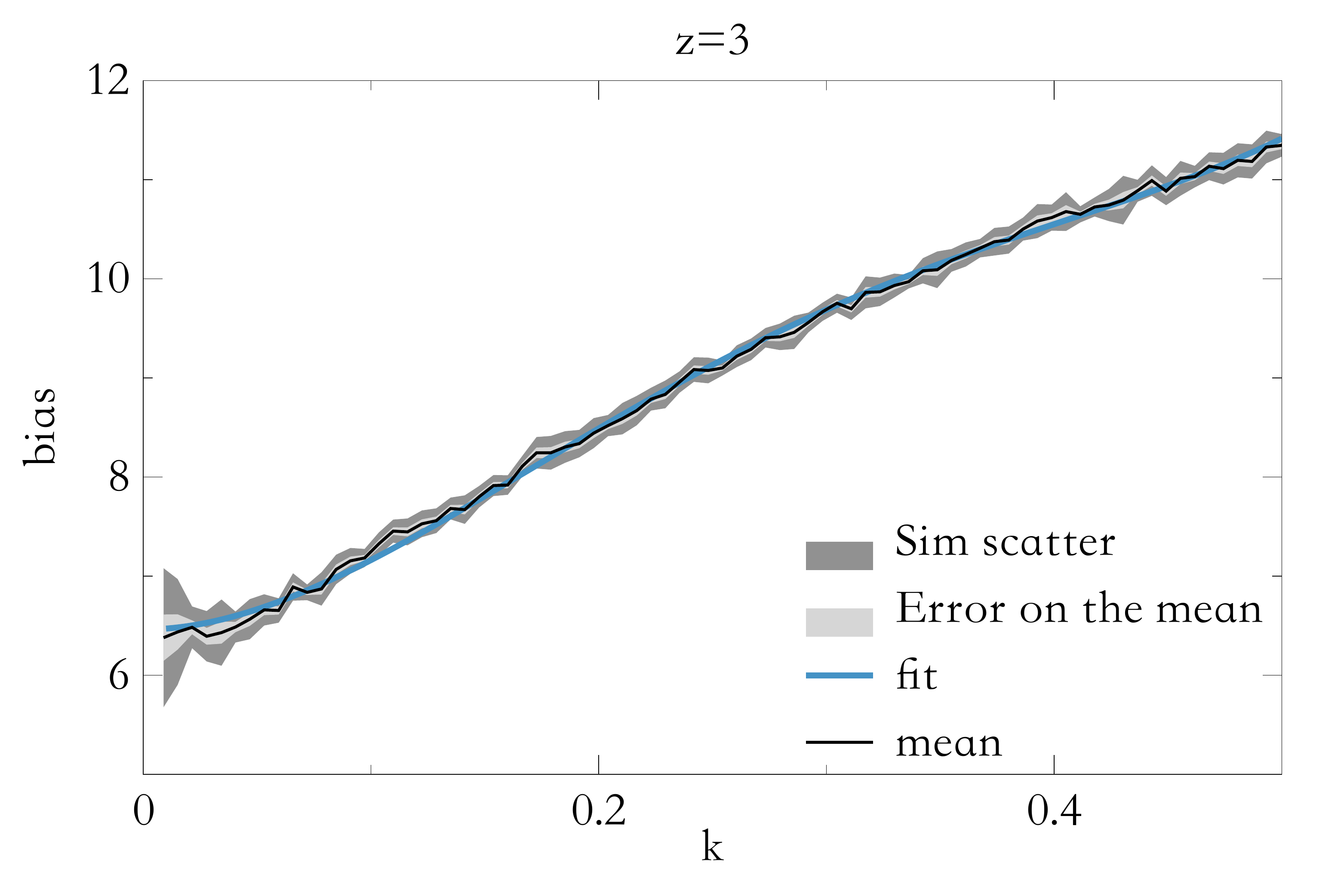}
\caption{Halo bias with respect to the cold dark matter $b_{\rm cc}$  as function of scale and  redshift. We show the mean of the 9 simulations (black jagged line), the error estimated as   {\it rms} among the simulations (dark gray band) and the error on the mean (light gray band). The smooth (blue) thick line is the employed smooth fit, see Table~\ref{tab:fitbcc}.}
\label{fig:bias_appendix}
\end{figure*}

\vspace{-0.2in}

\bibliography{Neutrino_bias}

\label{lastpage}
\end{document}